\documentclass[twocolumn]{aastex631}

\usepackage{amsmath}

\begin{document}
%\linenumbers
\title{An SDSS-V 3D Tomographic Na\,I Map of the ISM: An Initial Study Towards the Smith Cloud}

\author[0009-0001-5214-6330]{Timothy McQuaid}
\affiliation{Department of Astronomy, New Mexico State University, Las Cruces, NM 88003, USA}

\author[0000-0002-1979-2197]{Joseph N. Burchett}
\affiliation{Department of Astronomy, New Mexico State University, Las Cruces, NM 88003, USA}

\author[0000-0001-6248-1864]{Kate H. R. Rubin}
\affiliation{Department of Astronomy, San Diego State University, San Diego, CA 92182, USA}

\author[0000-0002-6050-2008]{Felix J. Lockman}
\affiliation{Green Bank Observatory, National radio Astronomy Observatory, Green Bank, WV 24944, USA}

\author[0000-0002-6561-9002]{Andrew K. Saydjari}
\affiliation{Department of Astrophysical Sciences, Princeton University, Princeton, NJ 08544 USA}

\author{Philipp Richter}
\affiliation{Institute für Physik und Astronomie, University of Potsdam, Karl Liebknecht Straße 24/25, 14476 Potsdam, Germany}

\author[0000-0003-0724-4115]{Andrew J. Fox}
\affiliation{AURA for ESA, Space Telescope Science Institute, 3700 San Martin Drive, Baltimore, MD 21218, USA}

\author[0000-0002-1793-3689]{David L. Nidever}
\affiliation{Department of Physics, Montana State University, Bozeman, MT 59717, USA}

\author{Jos\'e G. Fern\'andez-Trincado}
\affiliation{Universidad Cat\'olica del Norte, N\'ucleo UCN en Arqueolog\'ia Gal\'actica - Inst. de Astronom\'ia, Av. Angamos 0610, Antofagasta, Chile}

\author[0000-0002-9771-9622]{Jon A. Holtzman}
\affiliation{Department of Astronomy, New Mexico State University, Las Cruces, NM 88003, USA}

\accepted{March 20, 2026}
\submitjournal{ApJ}

\begin{abstract}
High velocity clouds supply the Milky Way with gas that sustains star formation over cosmic timescales. Precise distance measurements are therefore essential to quantify their mass inflow rates and gauge their exact contribution to the Galaxy’s gas supply. We use a sample of 594 SDSS-V BOSS stellar spectra within 10$^{\circ}$ of the high-velocity Smith Cloud (SC) to trace Na\,I absorption and dust extinction as functions of distance. By fitting ISM-corrected MaStar templates to each spectrum, we isolate residual equivalent widths and extinction then compare trends in the SC region to a same-latitude control field. Stars beyond 1 kpc toward the SC exhibit a significant Na\,I equivalent width excess ($>$0.2 \AA, $>$3$\sigma$) relative to the control. Two-component linear fits of Na\,I equivalent width and $A_{V}$ against both low and high-velocity H\,I column densities show that the low-velocity component is strongly correlated with both quantities, while the high-velocity term is marginally significant in extinction and Na\,I, consistent with a patchy, low dust-to-gas ratio. Given that the excess Na\,I begins at distances $<2$ kpc uniquely in the direction of the Cloud, and previous estimates of the SC place it at 12.4 $\pm$ 1.3 kpc, further investigation of its distance is warranted.

%Correlations of Na\,I equivalent width and A$_{V}$ with H\,I column density for the high-velocity gas are weaker (Pearson-r$\sim$0.2) than for low-velocity ISM (r$\sim$0.5), consistent with a patchy, low dust-to-gas ratio. 

%Together, these findings support a closer distance to the Smith Cloud and reveal a patchy distribution of cold, dusty gas that may elude standard surveys.

\end{abstract}



\section{Introduction} \label{sec:intro}

Gas accretion serves as a galaxy's lifeline, playing a vital role in sustaining star formation over cosmic timescales. Without a continuous supply of infalling gas, galaxies would deplete their gas reservoirs relatively quickly, within several Gyr \citep{Larson:1980vn,Leroy+2008}.  Additional evidence of fresh gas infall arises from the chemical abundances of stars \citep{Tinsley:1974zl}.  High-velocity clouds (HVCs) represent a significant mode of gas accretion, serving as a critical component in replenishing the galactic gas supply \citep{Lehner+2011, Marinacci+2010}. HVCs are defined kinematically by their lack of co-rotation with the host galaxy and high velocity in the Local Standard of Rest (v$_{LSR}$); therefore, they can originate from diverse processes, including inflows, outflows, and galactic fountain mechanisms \citep{Putman+2012}. 

%find paper/cite on lack of gas, something like half of SF uses gas accreted from CGM in last ~~years 

In the Milky Way (MW), we have an outstanding laboratory for studying cold gas accretion in the Smith Cloud (SC), a high-velocity cloud (HVC) that is infalling towards and might already be interacting with the MW disk. Considerable efforts have been devoted to measuring its properties, including extensive 21~cm emission mapping, which has provided information on H\,I column densities and kinematic measurements \citep{Lockman+2008, Lockman+2023}. The Smith Cloud has an elongated shape that stretches across approximately 10 degrees on the sky in H\,I emission. However, the sensitivity limitations of 21~cm observations restrict detection to column densities of \( N(\mathrm{H~I}) \gtrsim 10^{17} \, \mathrm{cm^{-2}} \), likely leaving significant (particularly ionized) portions of the HVC invisible. Absorption studies involving ultraviolet and optical spectra of distant active galactic nuclei have offered key insights into the Smith Cloud’s overall extent, phase structure, and metallicity, revealing that it has roughly half-solar metallicity \citep{Fox+2016} and contains dust \citep{Vazquez+2025}.

A key parameter in determining the total mass of the Smith Cloud, and consequently its potential to fuel star formation in our Galaxy, is its distance, which remains one of the largest sources of uncertainty. The currently accepted distance of 12.4$\pm$1.3 kpc \citep{Lockman+2008} is derived from a combination of several methods. \citet{Lockman+2008} used a kinematic model based on a flat Galactic rotation curve and the cloud's velocity relative to the surrounding interstellar medium, which was assumed to be corotating with the Milky Way disk, to estimate a distance of 12.4$\pm$1.3 kpc. \citet{Putman+2003} used H$\alpha$ emission measurements combined with photoionization models to constrain the Smith Cloud's distance to a range of 1.2--13.4 kpc. While these methods are valuable, they rely on assumptions about the ionizing radiation field or Galactic rotation, making them inherently indirect.  

The only direct method for determining distances to HVCs remains stellar spectroscopy: absorption lines from species such as Na\,I or Ca\,II in the spectra of stars with known distances can confirm whether the star lies behind the cloud (absorption present) or in front of it (absorption absent) \citep{Prata+1967}. This technique was applied to the Smith Cloud by \citet{Wakker+2008}, who observed eight stars and detected significant absorption in Ca\,II toward one star; they estimated a distance of 9.8--15.1 kpc for the cloud. A summary of their observations is provided in Table \ref{tab:Wakker Sources}. However, this study was conducted before the advent of high-precision astrometric surveys such as Gaia \citep{Gaia+2016,Gaia+2023}. At the time, stellar distances were inferred by fitting stellar models to observational data, introducing substantial uncertainty. This involved estimating stellar parameters (temperature, surface gravity, and metallicity) from spectroscopy and photometry, then placing stars on theoretical isochrones to derive absolute magnitudes and distances.

While high-resolution absorption spectroscopy provides the best separation between individual absorption components associated with HVC velocities and those of stellar atmospheres, this method is observationally intensive and limited to small sample sizes, particularly for stars at the relevant distances. To address these limitations, this study leverages a large sample of stellar spectra from the SDSS-V Milky Way Mapper (MWM) survey, which provides BOSS  spectra for a significant number of stars covering the Smith Cloud across a broad distance range. We describe these data and our analysis of them in Section \ref{sec:Analysis}; we present the resulting measurements and relations in Section \ref{sec:Results}.  We discuss the implications of our results in Section \ref{sec:Discussion} and summarize the work in Section \ref{sec:Conclusion}.

\section{Analysis} \label{sec:Analysis}

\subsection{SDSS-V MWM}
Milky Way Mapper (MWM) is one of three programs in the fifth generation of the Sloan Digital Sky Survey \citep[SDSS-V;][]{Kollmeier+2019, Kollmeier+2025}. The all-sky MWM is currently obtaining over 5 million stellar spectra across the Milky Way using the BOSS spectrograph (optical; 3,600--10,400~\AA; \citealt{Smee+2013}) and the APOGEE spectrograph (near-infrared; 1.51--1.70~$\mu$m; \citep{Wilson+2019}). Observations in the northern hemisphere are conducted with the Sloan 2.5-m telescope at Apache Point Observatory \citep[APO;][]{Gunn+2006}, and those in the southern hemisphere are conducted with the DuPont 2.5-m telescope at Las Campanas Observatory \citep[LCO;][]{Bowen+1973}.

The BOSS spectrograph has a spectral resolution of \(R \sim 2000\), corresponding
to an instrumental velocity scale of \(\sim 150~\mathrm{km\,s^{-1}}\). In 
practice, the ability to distinguish absorption or emission features from one another depends on this 
instrumental scale together with the intrinsic line widths and the velocity 
separation between components, which typically must be \(\gtrsim 1\)–2 resolution 
elements (\(\gtrsim 150\)–\(300~\mathrm{km\,s^{-1}}\)) for reliable deblending. 
Of primary concern in this study, the Na\,I and Ca\,II absorption associated with HVC gas is intrinsically narrow, with 
Doppler parameters \(b \sim 2\)–\(8~\mathrm{km\,s^{-1}}\) (corresponding to
\(\mathrm{FWHM} \lesssim 12~\mathrm{km\,s^{-1}}\); \citealt{Welsh+2009,Richter+2005}). However, even the `high-velocity' material associated with the Smith Cloud generally lies within \(\lesssim 100~\mathrm{km\,s^{-1}}\) of stellar and ISM absorption features, well below the BOSS resolution; thus, the features blend together. Consequently, BOSS spectra do not permit clean separation of SC absorption from overlapping components, unlike high-resolution studies with \(\lesssim 10~\mathrm{km\,s^{-1}}\) precision. Therefore, we herein leverage the large statistical sample of available stellar sightlines, utilize known information about absorption from the stellar atmospheres themselves, and measure the residual absorption along lines of sight to all stars matching our selection criteria.

%The BOSS spectrograph possesses a spectral resolution of \( R \sim 2000 \), corresponding to a velocity resolution of $\sim$150 km s$^{-1}$. This is significantly lower than the $<$10 km s$^{-1}$ resolution achieved in most high-resolution studies of HVC distances, thus the high velocity absorption component from the SC cannot be fully resolved from stellar and ISM absorption lines. We instead adopt a stellar modeling approach. By subtracting the modeled stellar spectrum, we isolate residual absorption associated with the interstellar medium (ISM) and high-velocity gas.

We source our stellar samples from the daily BOSS summary file within the SDSS Science Archive Server current to MJD 61016 and processed with the \texttt{v6.2.0} \texttt{idlspec2d} pipeline version \citep{Bolton+2012}. We first remove galaxies and active galactic nuclei using the \texttt{CLASS} keyword provided by the pipeline. We further exclude white dwarfs, also provided by \texttt{idlspec2d}, as these stellar types are poorly sampled by our stellar models and generally result in poor fits. Stellar parameters $T_{\mathrm{eff}}$, $\log g$, and [Fe/H] are provided by template fitting with \texttt{pyXCSAO} \citep{Kounkel+2022}, the Python
implementation of the \texttt{XCSAO} cross-correlation algorithm \citep{Kurtz+1992}. Because \texttt{pyXCSAO} determines these parameters by maximizing the cross-correlation over wide spectral regions, the resulting
matches are governed by broad continuum shape and global metal-line structure
rather than the narrow Na\,I D or Ca\,II H\&K absorption features produced
by the interstellar medium. As a result, the presence of ISM absorption in the
BOSS spectra does not systematically bias the derived metallicities.
To ensure high data quality, we apply a signal-to-noise ratio (SNR) cut, requiring that \texttt{SN\_MEDIAN\_ALL}, the median SNR over all valid pixels, is greater than 20. 

We adopt stellar distance estimates from \citet{BailerJones+2021}, who provide both geometric and photogeometric values derived from Gaia EDR3 parallaxes \citep{EDR3+2021}. Distances are inferred probabilistically using the parallax likelihood together with a direction-dependent prior based on the GeDR3 mock catalog \citep{Rybizki+2020}, a Besançon-like Galactic model \citep{Robin+2003} constructed with PARSEC evolutionary tracks \citep{Bressan+2012}. The photogeometric estimates additionally incorporate stellar color and apparent magnitude to restrict the range of plausible absolute magnitudes. For each star, we use the photogeometric distance along with the 16th and 84th percentile bounds of the posterior distribution. We also apply a standard Gaia parallax quality cut requiring a Renormalized Unit Weight Error (\texttt{ruwe}) $<$ 1.4, which removes stars with unreliable astrometry such as unresolved binaries. After applying these selection criteria, our final sample consists of 1,718,030 stars covering the entire sky.

\begin{figure}
    \centering
    \includegraphics[width=\linewidth]{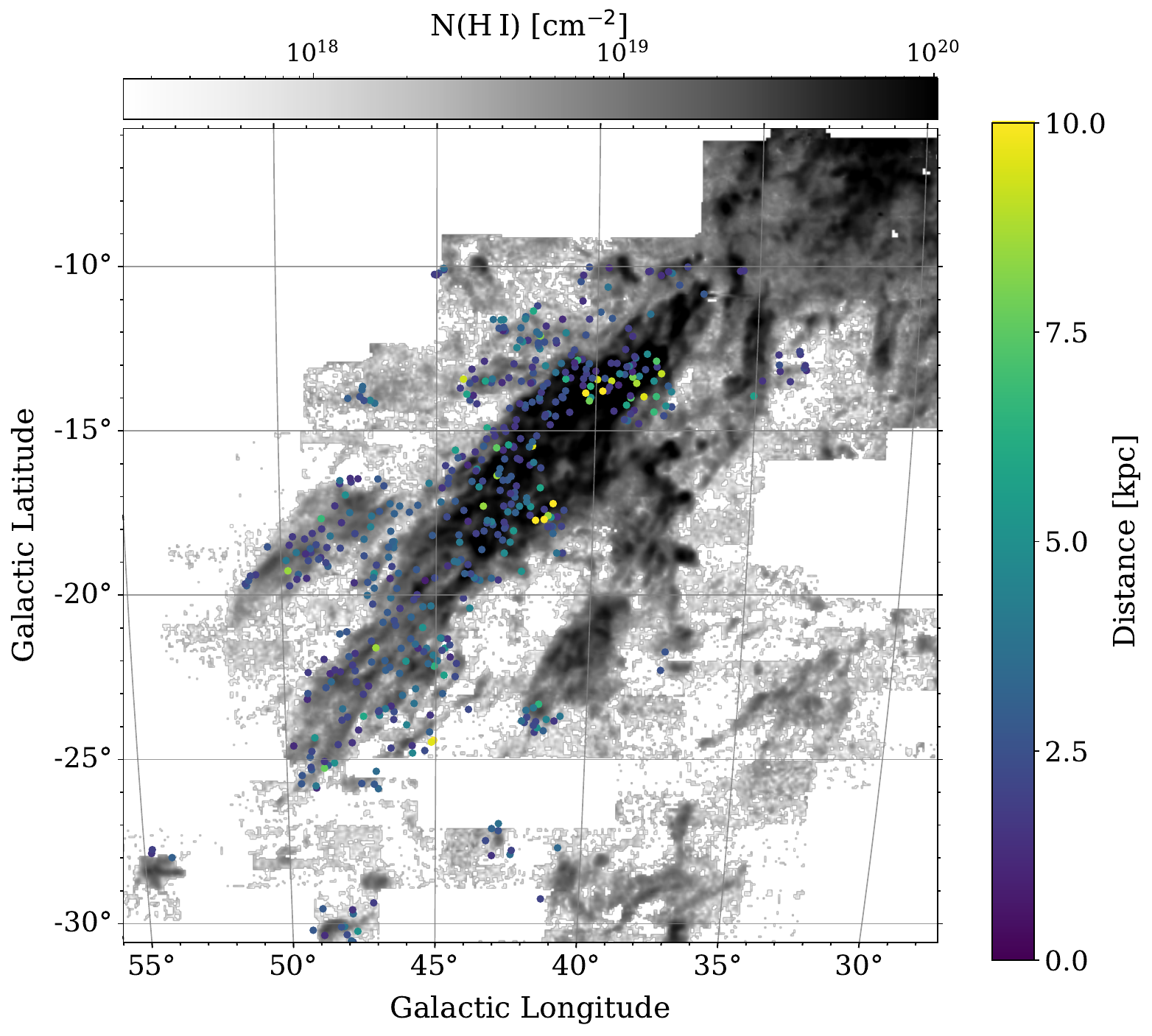}
    \caption{H\,I map of the main body of the Smith Cloud at $(l,b)=(43^\circ,-18^\circ)$, integrated over $v_{\rm LSR}=80$--150 km s$^{-1}$ from the Green Bank Telescope (private comm.). The sources coincident with strong 21-cm emission (594 BOSS sightlines), are marked and colored by distance.}

    \label{fig:SC_map}
\end{figure}

We additionally make use of a 21 cm H\,I emission datacube from the Green Bank Telescope (GBT) to trace the high-velocity neutral gas associated with the SC. These data are used both to map the SC’s spatial extent and to compare H\,I column densities with absorption and extinction measurements. While recent work by \citet{Lockman+2023} shows that the SC extends up to 40$^\circ$ in length, the GBT map used here covers a $\sim$20$^\circ$ region encompassing the main body of the SC. Figure~\ref{fig:SC_map} shows the resulting H\,I map alongside the positions and distances of our fitted MWM stellar sightlines in this region.

\subsection{Empirical Stellar Templates} \label{sec:stellar models}
Low resolution (R~$\lesssim$~2000) empirical stellar templates typically suffer from contamination from MW ISM absorption in key transitions such as Na\,I and Ca\,II. When normalizing a stellar spectrum by a best fit to obtain estimates of absorption, this contribution from the ISM in the template can lead to absorption attributed to interstellar gas being underestimated \citep{Rubin+2025}. 

We address this by adopting stellar templates that have been corrected for ISM
absorption in the Na\,I D and Ca\,II H \& K lines. These templates are drawn
from the Mapping Nearby Galaxies at Apache Point Observatory (MaNGA) Stellar
Library (MaStar), an SDSS-IV initiative using the BOSS spectrograph, and
comprise the largest empirical stellar library to date by an order of magnitude
\citep{Yan+2019, Abdurrouf+2022}. We use the version of this library in which the
contributions of Na\,I D and Ca\,II from the ISM are modeled as a function of
stellar distance, Galactic latitude, and dust reddening along the line of sight
\citep{Rubin+2025}. In this framework, the ISM corrections to Na\,I and Ca\,II
are performed independently; therefore, we construct from it two model grids for stellar fitting: one in which
all Na\,I-corrected spectral regions are valid, and one in which all
Ca\,II-corrected regions are valid. This ensures that each absorption line is
modeled using the subset of templates for which that specific ISM correction is
reliable.

\subsection{Fitting the Stellar Spectra} \label{sec:stellar fitting}

\begin{figure*}
    \centering
    \includegraphics[width=0.8\linewidth]{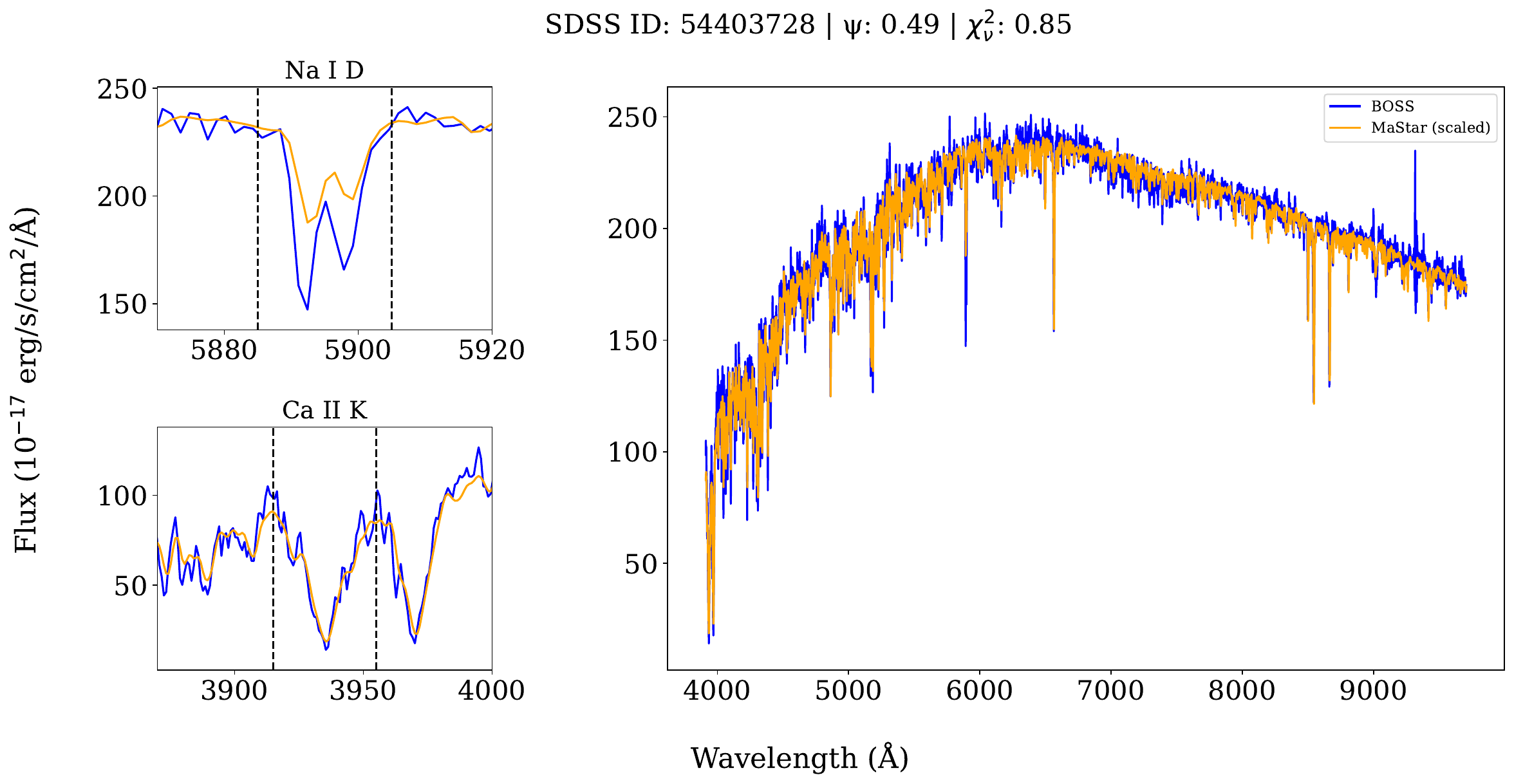}
    \caption{Example best-fit MaStar template to an observed MWM stellar spectrum. The large panel on the right shows the full-spectrum comparison, with the observed BOSS spectrum in blue and the best-fit, reddened, and scaled MaStar template in orange. The two subpanels on the left zoom in on key absorption features: the Na\,I~D doublet (top) and the Ca\,II~K region (bottom), again showing the BOSS spectrum in blue and the MaStar template in orange. The vertical black dashed lines in the Ca\,II~K panel indicate the window used for equivalent-width estimation of the Ca\,II~K line.}
    \label{fig:fitted_spectra}
\end{figure*}

\begin{figure}
    \centering
    \includegraphics[width=\linewidth]{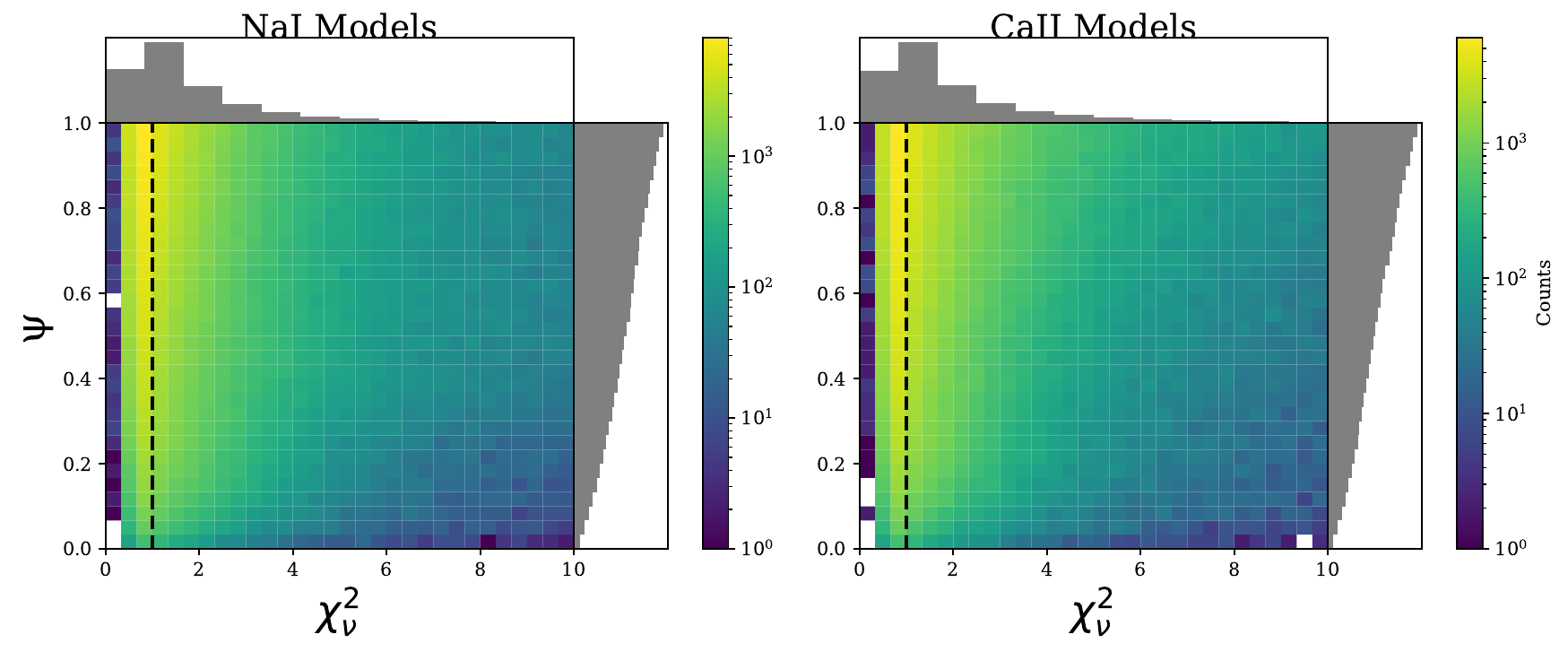}
    \caption{Distribution of $\chi^2_\nu$ versus $\psi$, as defined in Eq \ref{psi_eq}, for the fitted stellar sample. The left panel shows results for stars fitted using Na\,I-corrected MaStar templates, while the right panel shows the same for Ca\,II-corrected templates. Both samples exhibit a concentration towards $\chi^2_\nu\sim1$ and a modest preference towards $\psi\sim 1$, the upper bound allowed by our fitting procedure. The vertical dashed line denotes the $\chi^{2}_{\nu}$ cutoff of 1.}
    \label{fig:chi2_phi}
\end{figure}

To fit the stellar models, we first perform a grid search to match the values for
T$_{\mathrm{eff}}$, log\,\emph{g}, and [Fe/H] provided by the BOSS pipeline via
\texttt{pyXCSAO} to those of the MaStar library, and we identify template spectra
within close proximity in parameter space according to the following metric:

\begin{equation}\label{psi_eq}
\begin{aligned}
\psi(i,j) &=
\left(\frac{\log g_{i} - \log g_{j}}{\sigma_{\log g_{i,j}}}\right)^{2}
+\left(\frac{T_{\mathrm{eff},i} - T_{\mathrm{eff},j}}{\sigma_{T_{\mathrm{eff},i,j}}}\right)^{2} \\
&\quad
+\left(\frac{[\mathrm{Fe}/\mathrm{H}]_{i} - [\mathrm{Fe}/\mathrm{H}]_{j}}{\sigma_{[\mathrm{Fe}/\mathrm{H}]_{i,j}}}\right)^{2}
\end{aligned}
\end{equation}

where the quantities with subscript \textit{i} denote properties of the star being fitted and those denoted by \textit{j} correspond to the stellar model; $\sigma$ is the uncertainty of the quantity for both stars (\emph{i},\emph{j}) added in quadrature. We fit all MaStar spectra to a BOSS spectrum within $\psi < 1$, a broad cut that increases the pool of candidate templates and mitigates potential biases from the pipeline stellar parameters. The observed spectrum is first shifted to the stellar rest frame using the BOSS pipeline radial velocity. For each template, we apply the \citet{Fitzpatrick+1999} dust extinction law and determine the best-fit extinction $A_V$ by minimizing the residuals between the reddened template and the BOSS spectrum with a bounded least-squares optimization. We simultaneously fit a scalar normalization constant to account for distance. We repeat this for all templates within $\psi < 1$ for a single BOSS spectrum, and adopt the fit with the lowest $\chi^{2}_{\nu}$ as the best fit. A typical fit with $\chi^{2}_{\nu} \sim 1$ is illustrated in Figure~\ref{fig:fitted_spectra}. For analysis, we impose a $\chi^{2}_{\nu}$ cutoff of 1, which constitutes 26\% and 23\% of the fits in the Na\,I model and Ca\,II model samples, respectively. Figure \ref{fig:chi2_phi} shows a density map of the points in $\psi$ vs $\chi^{2}_{\nu}$. We find that the good fits span the allowable range of $\psi$ with little preference for higher or lower values.

\begin{figure*}
    \centering
        \includegraphics[width=0.8\linewidth]{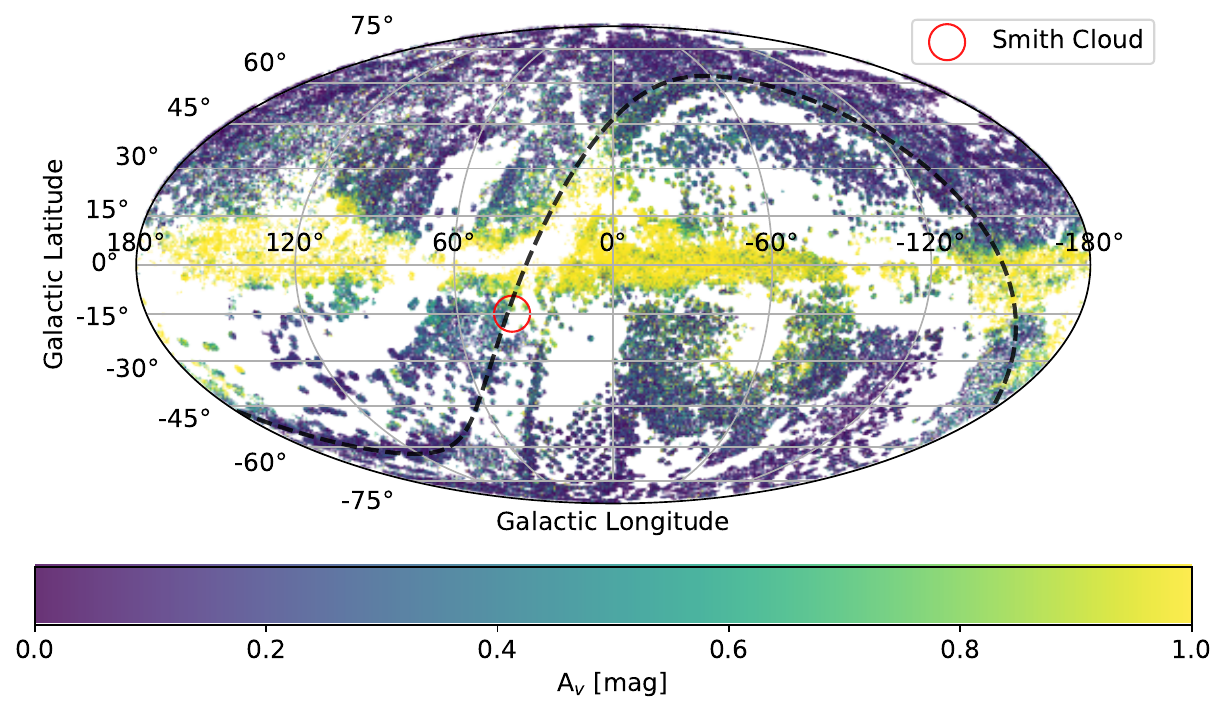}
    \caption{Dust extinction map of fitted SDSS-V BOSS stellar spectra (as of early July 2025), colormapped to the A$_{V}$ obtained from fitting stellar templates to the data using a \citet{Fitzpatrick+1999} extinction model. The black dashed line denotes a declination of 0$^{\circ}$, separating the APO (North) and LCO (South) observations. The plane of the Milky Way, as well as large dusty complexes at higher latitudes are clearly visible.}
    \label{fig:allsky_Av}
\end{figure*}

\begin{figure*}
    \centering
    \includegraphics[width=0.8\linewidth]{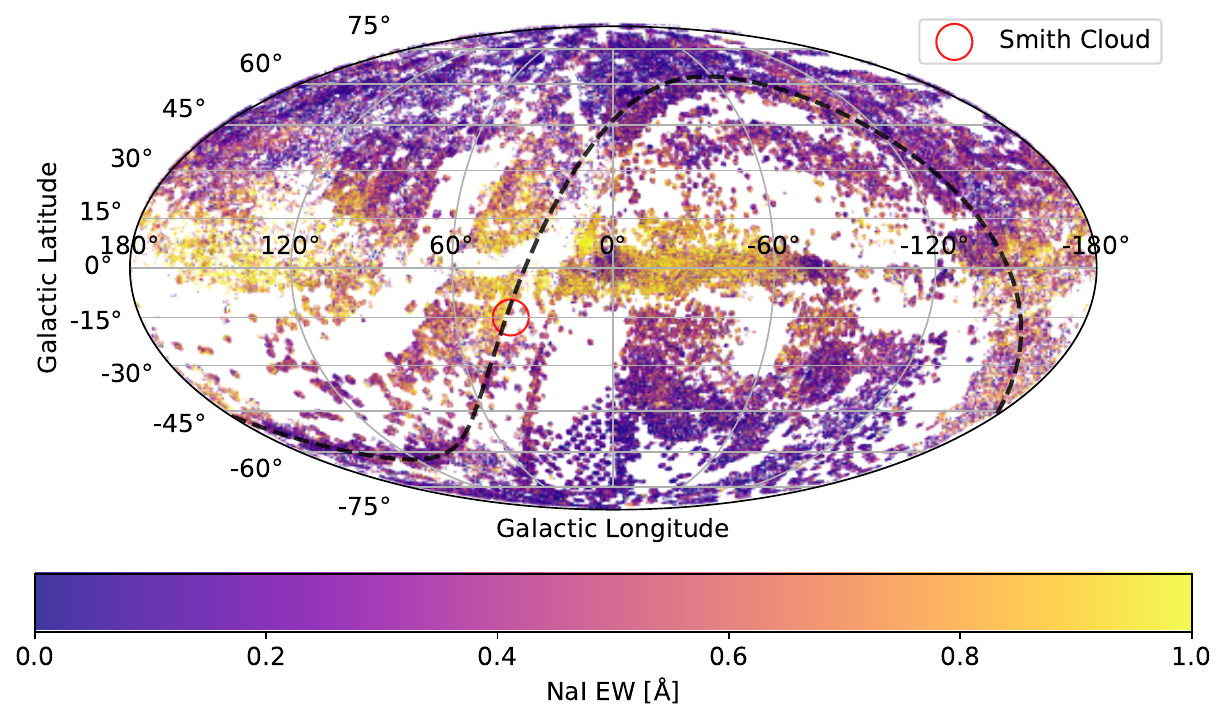}
    \caption{All-sky map of residual Na\,I absorption from SDSS-V BOSS spectra after fitting stellar templates and removing contributions from stellar atmospheres. The structure closely follows the $A_V$ map, consistent with Na\,I tracing cold neutral gas associated with dust.}
    \label{fig:allsky_NaI}
\end{figure*}

\begin{figure*}
    \centering
    \includegraphics[width=0.8\linewidth]{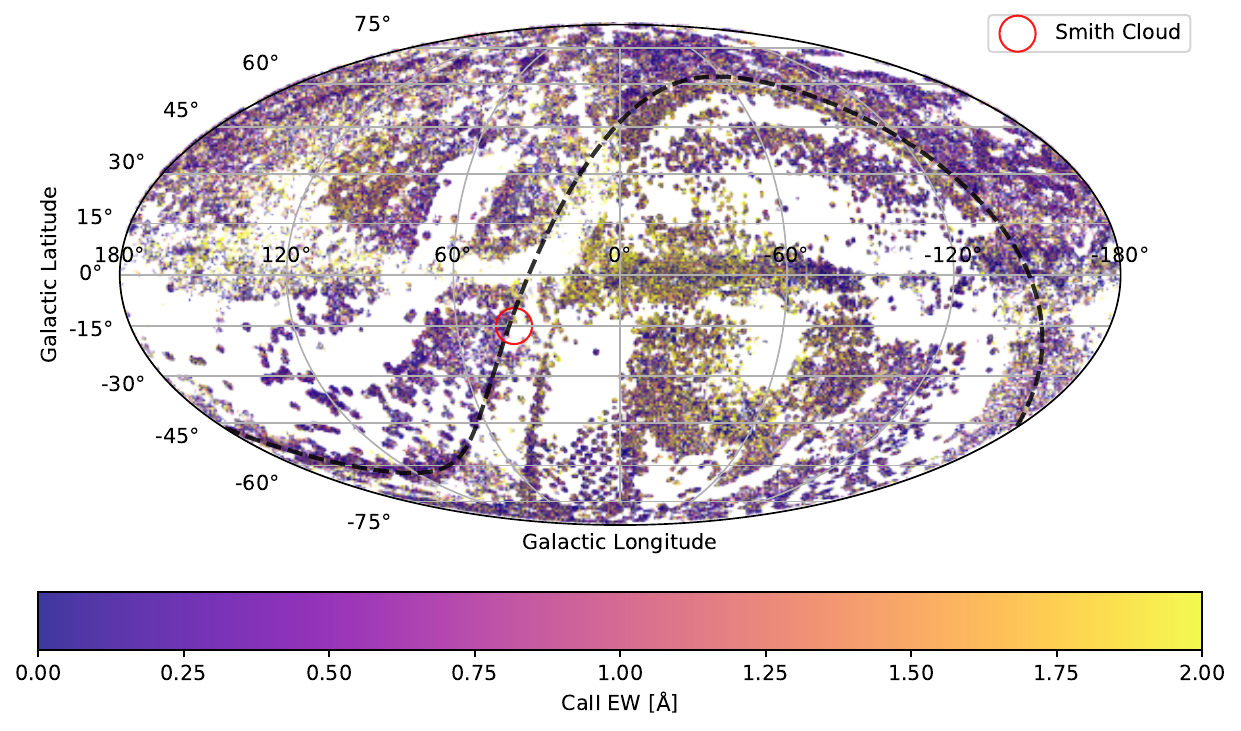}
    \caption{Similar to Figure \ref{fig:allsky_NaI} but for the Ca\,II K equivalent width.}
    \label{fig:allsky_CaII}
\end{figure*}

We compare the fitted extinction measurements to values from the 3D dust maps of \citet{Green+2019}, using the median stellar distances from \citet{BailerJones+2021}. We find a moderately close correlation between the two (Pearson correlation coefficient $r=0.72$) with an RMS scatter of 0.29 mag; this correlation vanishes when the stellar distances are randomized. A small subset of nearby stars ($\sim$2000) exhibit high fitted extinctions ($>1$ mag), while both the dust maps and $E(B-V)$-based estimates indicate negligible reddening along those sightlines. This discrepancy likely arises for two reasons. First, the 3D dust maps have angular resolutions of $\sim$3--14 arcmin \citep{Green+2019}, much larger than the PSF of the spectroscopic observations, and may therefore smooth over small-scale dust structure so that localized pockets of extinction are missed. Second, these stars lie close to our adopted SNR$>$20 threshold, introducing a selection effect: intrinsically similar stars with comparable reddening at larger distances would have lower SNR and fail our cut, biasing the sample toward detecting such reddened stars only at small distances.

To mitigate this incompleteness, we apply a completeness cut based on signal-to-noise. Specifically, we use distance as a proxy for apparent flux and rescale each star's observed spectrum assuming inverse-square dimming to determine the distance at which its SNR would fall below 30. Stars for which this occurs before reaching 1 kpc are excluded. This procedure does not modify the measured SNR of the observed spectrum; rather, it tests whether a star of the same intrinsic brightness and extinction would remain detectable over the multi-kiloparsec distance range relevant for our analysis. Applying this cut together with our $\chi^2_\nu$ threshold yields final samples of 192{,}396 stars fitted with Na\,I-corrected templates and 152{,}043 stars fitted with Ca\,II-corrected templates.

We define the Smith Cloud sample using GBT 21-cm data. Emission is integrated over LSR velocities of 80--150 km s$^{-1}$, and sightlines exceeding a column density threshold of $1.1 \times 10^{19}$ cm$^{-2}$ are identified. To account for lower-column-density material surrounding the main structure, we further include all sightlines lying on or within $0.5^\circ$ (10 pixels) of this threshold. The resulting Smith Cloud sample comprises 594 sightlines, shown in (Figure~\ref{fig:SC_map}).

We measure the excess equivalent widths (EWs) of the Ca\,II K and Na\,I D absorption attributable to non-stellar material (ISM and high-velocity gas) by comparing each observed BOSS spectrum, $F_{\mathrm{data}}(\lambda)$, to its best-fit MaStar template, $F_{\mathrm{model}}(\lambda)$. The MaStar template represents the stellar photospheric spectrum plus the smooth continuum. We define a continuum-normalized residual depth:
\begin{equation}
    D(\lambda) \;=\; 1 \;-\; \frac{F_{\mathrm{data}}(\lambda)}{F_{\mathrm{model}}(\lambda)} ,
\end{equation}
and compute the equivalent width of the non-stellar component within a fixed wavelength window $[\lambda_1,\lambda_2]$ as
\begin{equation}
    \mathrm{EW}_{\mathrm{ISM}}
    \;=\;
    \int_{\lambda_1}^{\lambda_2} D(\lambda)\, d\lambda
    \;=\;
    \int_{\lambda_1}^{\lambda_2}
      \left[1 - \frac{F_{\mathrm{data}}(\lambda)}{F_{\mathrm{model}}(\lambda)}\right]
      d\lambda .
\end{equation}
In practice, this integral is evaluated as a discrete sum over the BOSS wavelength grid. We adopt fixed windows of 3915--3955~\AA\ for Ca\,II\,K, 3955--3990~\AA\ for Ca\,II\,H, and 5885--5905~\AA\ for Na\,I\,D.

We also compute line-of-sight velocity centroids for both the Na\,I doublet and the Ca\,II H and K residual absorption profiles. Using the same residual depth $D(\lambda)$, we convert wavelength to velocity via
\begin{equation}
    v \;=\; c \left(\frac{\lambda}{\lambda_0} - 1\right) ,
\end{equation}
where $\lambda_0$ is the appropriate rest wavelength. For Na\,I we treat the blended D$_1$ and D$_2$ components using an oscillator-strength-weighted rest wavelength
\begin{equation}
    \lambda_{0,\mathrm{Na\,I}}
    \;=\;
    \frac{f_{D2}\,\lambda_{D2} + f_{D1}\,\lambda_{D1}}{f_{D2} + f_{D1}} ,
    \qquad
    f_{D2} = 0.641,\;\; f_{D1} = 0.320,
\end{equation}
while for Ca\,II we treat the H and K lines separately using their laboratory rest wavelengths. In all cases, the velocity centroid is given by the first moment of the residual absorption profile, evaluated numerically over the same fixed windows used for the EW measurements. Uncertainties on the EWs and centroids are propagated pixel-by-pixel from the flux residual uncertainties.

% We determine the equivalent widths (EWs) of the Ca\,II K and Na\,I D lines in both the BOSS and MaStar spectra. To estimate the continua, we first apply a median filter to suppress sharp absorption features. We then fit a third degree polynomial to the spectrum and apply a smoothing filter for additional refinement. We finally measure the equivalent widths using the \texttt{specutils} Python package \citep{Earl+2024} within fixed wavelength windows (3915--3955 \AA\ for Ca\,II, 5885--5905 \AA\ for Na\,I). The difference in EWs between the BOSS and MaStar spectra represents the residual absorption attributed to non-stellar components. We also measure the velocity centroid of the blended Na\,I doublet, weighting the contribution of each component by its oscillator strength (\(f_{D1}=0.320\), \(f_{D2}=0.641\)).
\section{Results} \label{sec:Results}

\subsection{All-sky Results}
To validate our fitting procedure and derived measurements, we examine the all-sky distribution of extinction and Na\,I residual absorption in galactic coordinates. The extinction map (Figure~\ref{fig:allsky_Av}) clearly traces known features of the Milky Way, including the galactic plane and high-latitude dust structures. The residual Na\,I absorption (Figure~\ref{fig:allsky_NaI}) exhibits similar features, reflecting the close association between cold neutral gas and dust \citep{Phillips+1984, Sembach+1994,Munari+1997,Welty+2006, Welty+2012,Poznanski+2012,Murga+2015}. 

% The Ca\,II residual absorption is comparatively uniform across the sky, exhibiting a much weaker dependence on Galactic latitude than seen for Na\,I. In contrast to the strong mid-plane concentration of Na\,I, the Ca\,II signal remains broadly distributed across both low and high latitudes, consistent with Ca\,II tracing warmer, more diffuse interstellar gas \citep{Phillips+1984, Sembach+1994, Munari+1997, Murga+2015,}.
In contrast, the Ca\,II distribution shows a clear dichotomy between the northern and southern skies: the measured equivalent widths from APO (north) are systematically smaller than those from LCO (south), with an apparent enhancement across the southern hemisphere. We note that Ca\,II~K lies near the blue edge of the BOSS wavelength coverage, where the S/N is universally lower than at redder wavelengths and data-reduction/calibration systematics may be more pronounced. The Ca\,II residuals are therefore more likely  to be susceptible to such effects. Because of this issue, we hereafter restrict our analysis to the Na\,I results, which provide a more consistent and reliable probe of both large-scale ISM structure and high-velocity cloud gas and defer further analysis employing the Ca\,II statistics until these systematics are better understood.

\subsection{Derived Properties vs. Distance}
\begin{figure*}
    \centering
    \includegraphics[width=1.0\linewidth]{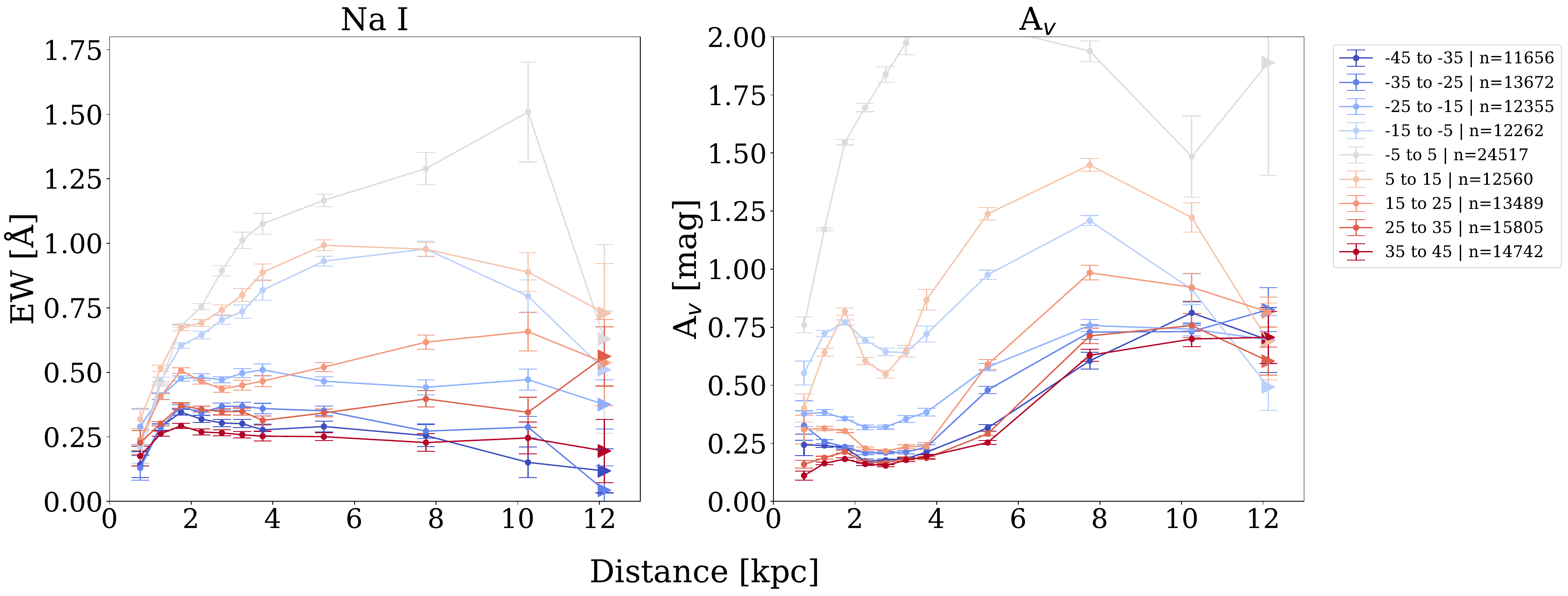}
        \caption{Median residual Na\,I equivalent width (left) and dust extinction $A_V$ (right) versus distance for stars in $10^\circ$ Galactic latitude bins. More faintly colored lines are at lower $|b|$ (closer to the Galactic plane). In the Na\,I panel, at high latitudes, the absorption quickly plateaus after the initial rise, whereas low-latitude sightlines continue to increase, reflecting the thickness of the disk. The $A_V$ panel reveals a pronounced early rise for low latitudes and a smoother ascent at higher latitudes.}
    \label{fig:latitudes}
\end{figure*}

We now examine how Na\,I and Ca\,II equivalent widths and dust extinction vary with distance in $10^\circ$ slices of Galactic latitude, grouping stars in distance bins of [0, 1, 2, 3, 4, 6.5, 9, 11.5, 14] kpc. As shown in Fig.~\ref{fig:latitudes}, Na\,I rises sharply within the first few kiloparsecs for low-$|b|$ sightlines, reflecting the dense cold gas in the Galactic plane. Because Na\,I predominantly traces cold, neutral gas with $T \lesssim 10^{3}$\,K \citep{Crawford+1992,Welty+1996,Puspitarini+2012}, this rapid increase at low $|b|$ is expected where sightlines intersect the coldest ISM phases, consistent with the strong galactic latitude dependence of Na\,I absorption \citep{Murga+2015}. In higher-latitude bins, Na\,I levels off beyond 4~kpc at EW values of $\sim$0.4\,\AA\ as sightlines exit the midplane. Low-latitude bins, in contrast, continue a gradual increase beyond 4~kpc and eventually reach EW values of 0.8 to 1.0\,\AA. Dust extinction, A$_{V}$, increases steeply at low-$|b|$ and low distance but increases more gradually at higher $|b|$. These distance- and latitude-dependent profiles trace the ISM’s stratified structure, capturing the concentration of cold gas and dust in the midplane.

%This distance-dependent trend in Na\,I is expected as a manifestation of the Routly–Spitzer effect: high-velocity clouds at larger Galactocentric distances tend to have enhanced gas-phase Ca\,II due to dust grain destruction and altered ionization relative to Na\,I. As stars beyond $\sim$6~kpc begin to intersect this halo gas, the Ca\,II equivalent width rises sharply, whereas Na\,I has already plateaued once the local cold gas is fully sampled \citep{Routly+1952, Bekhti+2012}.

\begin{figure*}
    \centering
    \includegraphics[width=1.0\linewidth]{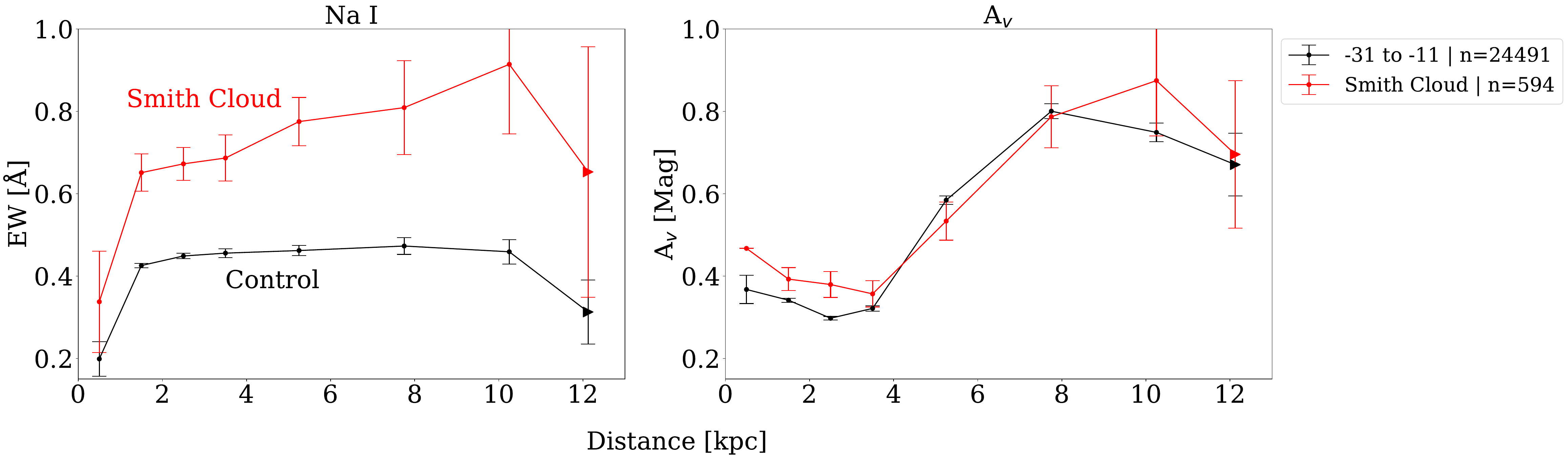}
    \caption{Similar to Figure \ref{fig:latitudes} but focusing on the comparison between the Smith Cloud (red) and same-latitude control sample (black). The Na\,I equivalent widths toward the Smith Cloud show an enhancement at distances $>$1 kpc but no significant difference in extinction compared to the control.}
    \label{fig:sc_control}
\end{figure*}

Applying this framework to the SC region, we compare the previously defined SC sample, consisting of 594 stellar sightlines lying on or near regions of strong H\,I emission, to a same-latitude control bin ($-31^\circ<b<-11^\circ$) in Figure~\ref{fig:sc_control}. The control sample includes 24,491 stellar sightlines spanning $-31^\circ<b<-11^\circ$ and all Galactic longitudes. This comparison reveals a pronounced Na\,I excess beyond 1~kpc in the Smith Cloud sightlines (220~m\AA, 4.80$\sigma$), which increases further at larger distances (310~m\AA, 4.30$\sigma$ between 4 and 6.5~kpc).

We do note that the nearest distance bin already exhibits a nominal Na\,I excess in the Smith Cloud sample relative to the control, although this difference is less than 1$\sigma$. To test whether the enhancement seen at larger distances could simply be inherited from this initial offset, we performed a differential equivalent-width analysis in which the EW in the lowest distance bin was subtracted from all subsequent bins for both the Smith Cloud and control samples. After this baseline subtraction, the Smith Cloud sightlines still show a positive Na\,I excess relative to the control beyond 1~kpc, with a nominal amplitude of $\sim$0.1~\AA. This is smaller than the $\sim$0.2~\AA\ excess inferred from the original, non-differential comparison, as expected because the differential analysis removes the $\sim$0.1~\AA\ offset already present in the nearest bin. The differential signal remains within 1$\sigma$ in all bins out to 10~kpc, reflecting both the reduced amplitude of the excess and the larger propagated uncertainties introduced by the differencing. Although the differential signal has a smaller amplitude than the nominal excess seen at larger distances in the original comparison, it nevertheless shows a nominal Na\,I excess in the Smith Cloud sightlines relative to the matched control sample beyond 1~kpc.

In contrast to Na\,I, the extinction behavior in the Smith Cloud sample closely follows that of the control sightlines. Aside from a slight excess at distances $\lesssim 4$~kpc, the Smith Cloud and control samples are statistically consistent across the full distance range. A slight decline in the Smith Cloud extinction within the nearest distance bins is visible; however, this feature is likely driven by the smaller number of sightlines at low distances combined with underestimated extinction uncertainties. The extinction fitting procedure yields formally small errors, reflecting only the local curvature of the least-squares solution rather than the full systematic uncertainty associated with continuum mismatches, stellar template imperfections, and dust law assumptions. As a result, we do not interpret this low-distance decrease in extinction as physical or significant.

\subsection{Validation of Na\,I excess relative to N(H\,I)}
To further scrutinize this result as potentially arising from an anomalously high gas column density in the direction of the Smith Cloud, we investigate the spatial variation of N(H\,I) and Na\,I EW within the same-latitude control bin.  We subdivide the stellar sightlines into a set of localized fields and examine how the foreground gas content and absorption vary across the sky at constant latitude by repeating the analysis in $10^\circ$ radius apertures spaced every $20^\circ$ in longitude at $b = -18^\circ$. Using the all-sky H\,I column density map from \citet{Bekhti+2016}, we measure the total integrated $N$(H\,I) in each subregion over all velocities. Across these fields, the median H\,I column density is $9.25 \times 10^{20}$ cm$^{-2}$, with a standard deviation of $2.60 \times 10^{20}$ cm$^{-2}$. The value in the Smith Cloud region is $8.55 \times 10^{20}$ cm$^{-2}$, or approximately 0.25$\sigma$ below the median; i.e., the neutral gas column density is not anomalously large in the Smith Cloud direction.

\begin{figure}
    \centering
    \includegraphics[width=0.9\linewidth]{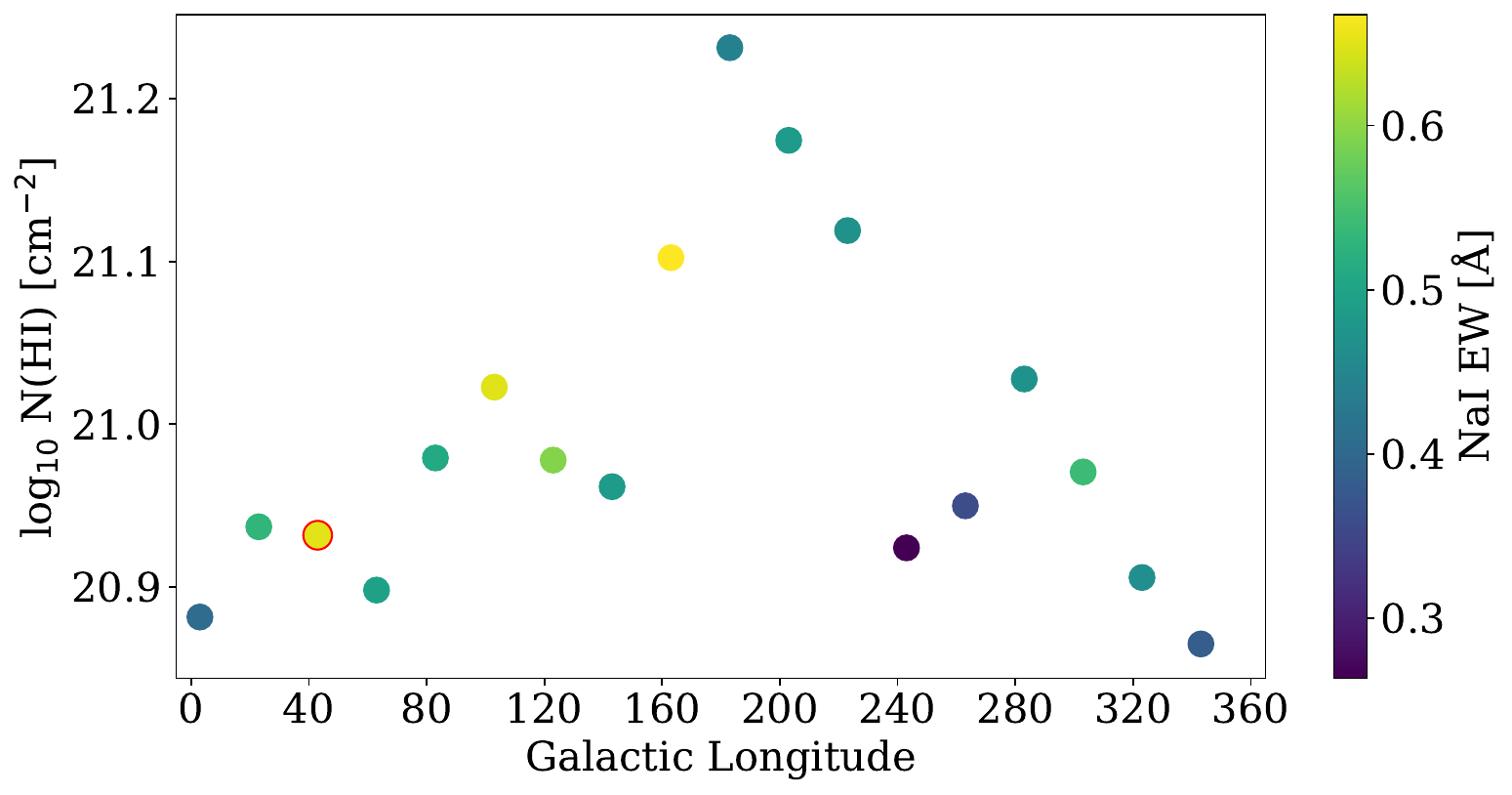}
    \caption{Average H\,I column density within $10^{\circ}$ radius apertures at fixed Galactic latitude with varying Galactic longitude. Points are color-coded by the median Na\,I D equivalent width measured in stellar sightlines at distances of 1--2 kpc for stars within each aperture; the red-outlined marker and error bar mark the bin corresponding to the Smith Cloud. Although the Smith Cloud region has a slightly lower-than-average $N$(H\,I), it shows comparatively stronger Na\,I absorption relative to neighboring longitudes.}
    
    \label{fig:NaI_NHI}
\end{figure}

We also measure the Na\,I D equivalent width profiles as a function of distance for each subregion. In the 1--2 kpc bin, the SC region exhibits the third-highest median Na\,I absorption, exceeded only by 2 subregions that both have greater H\,I column densities, at $l\sim100^{\circ}$ and $l\sim170^{\circ}$ as shown in Figure~\ref{fig:NaI_NHI}. The Smith Cloud region therefore appears somewhat distinct in exhibiting elevated Na\,I absorption at intermediate distances relative to its below-average H\,I content.

\begin{figure}
    \centering
    \includegraphics[width=1.0\linewidth]{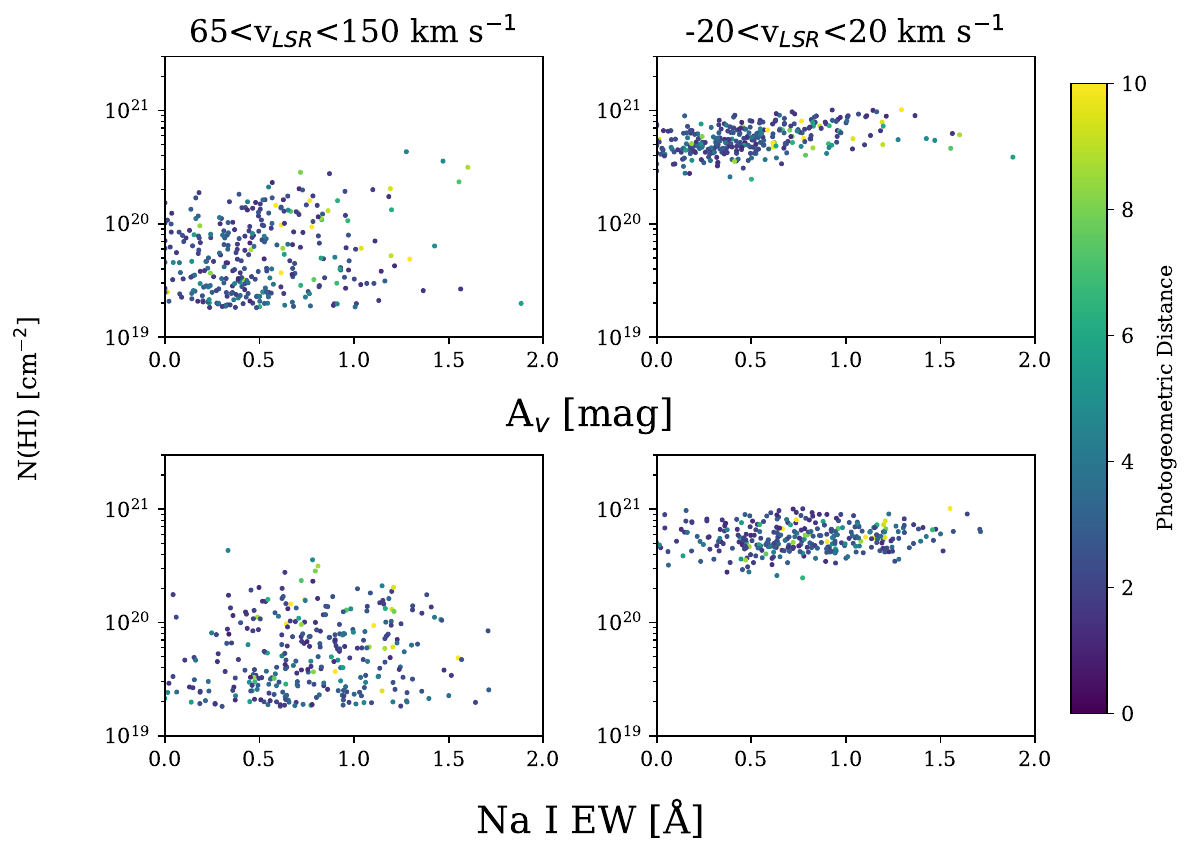}
    \caption{Correlations between neutral hydrogen column density N(H\,I) and extinction A$_{V}$ (top row) and Na\,I equivalent width (bottom row), for high-velocity gas (left column, $65 < v_{\mathrm{LSR}} < 150~\mathrm{km\,s^{-1}}$) and low-velocity gas (right column, $-30 < v_{\mathrm{LSR}} < 30~\mathrm{km\,s^{-1}}$) within 10$^\circ$ of the Smith Cloud. Points are colored by stellar distance.} 
    \label{fig:HI_Av}
\end{figure}

In light of this lack of correlation between H\,I and Na\,I across galactic longitudes, we next examine whether extinction and Na\,I absorption toward the Smith Cloud correlate with the low- and high-velocity components of $N(\mathrm{H\,I})$. Figure~\ref{fig:HI_Av} shows Na\,I equivalent width and extinction as a function of low-velocity ($-20 < v_{\mathrm{LSR}} < 20$ km s$^{-1}$) and high-velocity ($65 < v_{\mathrm{LSR}} < 150$ km s$^{-1}$) integrated H\,I column density from the GBT along each stellar line of sight. To account for the strong foreground contribution, we test a two-component linear model of the form
\begin{equation}
    Y = C + A \, N_{\mathrm{HI,low}} + B \, N_{\mathrm{HI,high}},
\end{equation}
where $Y$ denotes the dependent variable (either $A_V$ or Na\,I equivalent
width), $N_{\mathrm{HI,low}}$ and $N_{\mathrm{HI,high}}$ are the low- and
high-velocity H\,I column densities, and $C$, $A$, and $B$ are the best-fit
coefficients.

We estimate $C$, $A$, and $B$ using a weighted linear least-squares (WLS) fit. Specifically, we adopt the
\texttt{statsmodels} \citep{Statsmodels} WLS
implementation, with design matrix
$X = [1,\, N_{\mathrm{HI,low}},\, N_{\mathrm{HI,high}}]$ and weights
$w_i = 1/\sigma_{Y,i}^2$ based on the uncertainties in $Y$ (either
Na\,I EW or $A_V$). Uncertainties in the H\,I column densities are neglected in
the fit and are assumed to be minor compared to the errors on $Y$. To
avoid fitting sources below the H\,I detection limit of the GBT data, we remove
sources with high-velocity H\,I column densities below
$1.8 \times 10^{19}\,\mathrm{cm}^{-2}$ (3$\sigma$ above the high-velocity
background), leaving 313 out of 594 sources in the regression.

In the extinction regression, the low-velocity term is highly significant, $A = (7.6 \pm 0.4)\times10^{-22}$ mag\,cm$^{2}$ ($p \ll 0.001$), consistent with previous measurements of the galactic dust-to-gas ratio \citep{Bohlin+1978, Diplas+1994, Liszt+2014, Lenz+2017}. The fitted slope for the high-velocity component is comparable in magnitude, $B = (9.5 \pm 2.9) \times 10^{-22}$~mag~cm$^2$ ($p = 0.001$). To verify that this correlation is independent of the low-velocity ISM, we computed the residuals after removing the linear dependence on low-velocity HI from both $A_v$ and the high-velocity HI column density. These residuals still exhibit a significant positive correlation (Pearson $r = 0.204$, $p = 0.0003$), confirming that high-velocity gas contributes independently to the extinction signal, though the larger uncertainty suggests that any associated extinction from the Smith Cloud itself remains weak. Overall, the extinction signal is dominated by the foreground ISM, with at most a minor additional contribution from the high-velocity gas, consistent with the Cloud's low dust content \citep{Minter+2024}.

For the Na\,I regression, the low-velocity term 
($X_{1} = (1.13 \pm 0.06)\times10^{-21}$ \AA\,cm$^{2}$, $p \ll 0.001$) is significant, albeit shallow in slope. \citet{Ferlet+1985} measured $N(\mathrm{Na\,I})/N(\mathrm{H\,I})$ along stellar sightlines in the Milky Way, parameterizing their results as:
\begin{equation}
\log N(\mathrm{Na\,I}) = 1.04 \,[\log N(\mathrm{H\,I}+ \mathrm{H}_{2})] - 9.09.
\end{equation}
Converting this to units of \AA\,cm$^{2}$, while assuming a linear curve of growth with 
$N(\mathrm{H\,I}) \gg N(\mathrm{H}_{2})$, we find their slope to be within 12--20\% of ours 
over the same $N(\mathrm{H\,I})$ range as our low-velocity sample ($10^{20}$--$10^{21}$ cm$^{-2}$).

The high-velocity coefficient is $B = (1.02 \pm 0.38)\times10^{-21}$ \AA\,cm$^{2}$ ($p = 0.008$), indicating a formally significant but weak positive association between Na\,I and high-velocity H\,I. This coefficient is an order of magnitude less significant than the low-velocity term and is strongly affected by covariance between the two components, as reflected by the large condition number of the fit (2.32$\times$10$^{21}$). As a result, the high-velocity contribution is not robustly constrained and, even if real, represents a minor effect compared to the dominant foreground ISM.

Indeed, previous studies have shown that Na\,I does not correlate reliably with high-velocity H\,I. \citet{Wakker+2001} found that Na\,I/H\,I ratios in HVCs vary by more than an order of magnitude within a single cloud. \citet{Bekhti+2012} detected Na\,I in only $\sim$20--35\% of high-N(H\,I) halo sightlines compared to $\sim$40--55\% in Ca\,II. Our own test comparing integrated H\,I with median Na\,I absorption across galactic latitudes (Figure \ref{fig:NaI_NHI}) likewise revealed little correlation between the strengths of H\,I and Na\,I, even though the signal was dominated by low-velocity gas. If the relationship is already weak in the dense ISM, it is likely to be even less reliable in more diffuse high-velocity gas. In this context, our non-significant regression coefficient is in line with expectations, though the large number of sightlines toward the Smith Cloud may still permit a statistical detection of excess Na\,I relative to the control field.

\section{Discussion}
\label{sec:Discussion}

Our large-sample statistical measurements of Na\,I absorption toward stars within 10$^\circ$ of the SC reveal a clear enhancement at distances $>$1 kpc, consistent with the presence of additional cold gas along those sightlines relative to those at similar latitude but different longitudes. 

\subsection{High-Resolution Na\,I Spectroscopy and Covering Fraction}
To further corroborate this result, we conducted a follow-up program using the echelle spectrograph (ARCES; R$\sim$31,500) at Apache Point Observatory. This dataset includes 33 stars spanning distances from 2 to 16 kpc, each observed with SNR$>$15 at 5890 \AA. 

\begin{figure*}
    \centering
    \includegraphics[width=0.7\linewidth]{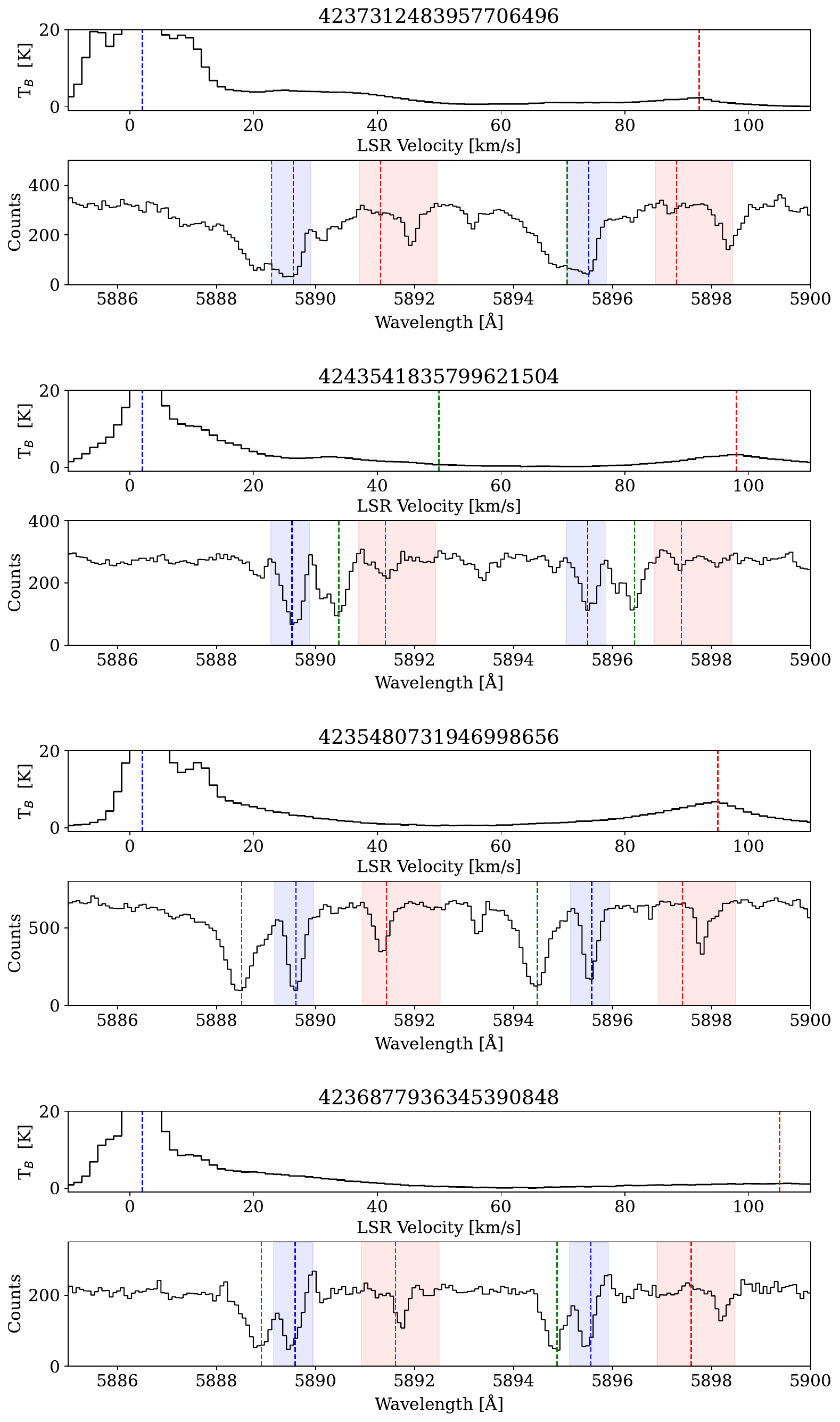}
    \caption{High-resolution ARCES spectra of four stars observed toward the Smith Cloud, with $>$3$\sigma$ detections in Na\,I D$_{2}$. Each is shown with an associated 21-cm spectrum extracted along the same sightline. For each star, the lower panel shows the optical spectrum in the region of the Na\,I D doublet, while the upper inset displays the corresponding H\,I brightness temperature as a function of LSR velocity. Vertical dashed lines indicate the velocities of the local interstellar medium (blue), the stellar radial velocity (green), and the Smith Cloud (red). Lightly shaded regions mark the velocity intervals of the equivalent width estimation: $-20$ to $+20$ km s$^{-1}$ (low velocity ISM, blue) and $70$ to $150$ km s$^{-1}$ (Smith Cloud, red).}
    \label{fig:echelle}
\end{figure*}

As these are distant F, G, and K stars, the SNR in the Ca\,II H/K region was too low to be usable. We searched each spectrum for absorption features, comparing the velocities of Na\,I components to both HI emission along the line of sight and the star’s own radial velocity. Typical sightlines exhibit a strong interstellar component near 0 km s$^{-1}$ and stellar component at the star’s radial velocity. We measure equivalent widths at low velocity (-20 -- 20 km s$^{-1}$ LSR) and at high velocity (70 -- 150 km s$^{-1}$ LSR); the results are compiled in Table \ref{tab:apo_targeted}. Four spectra at photogeometric distances between 3.8 and 16.7 kpc exhibit high-velocity Na\,I D$_2$ absorption at the 3$\sigma$ level; these are shown in Figure \ref{fig:echelle} and noted in Table \ref{tab:apo_targeted}. Two of these also show significant D$_1$ absorption, though the D$_1$ features appear slightly offset in velocity relative to their corresponding D$_2$ components. We checked for nearby stellar transitions from other species in the vicinity of the Na\,I lines using the Vienna Atomic Line Database (VALD; \citealt{Piskunov+1995}). Although these neighboring features have considerably smaller transition strengths than Na\,I D, several fall near our equivalent-width integration regions and coincide in stellar velocity with nearby absorption features, making them plausible contributors to some of the apparent high-velocity absorption. A definitive separation of Na\,I absorption from neighboring stellar lines will require more detailed spectral modeling, including the relevant species and abundances, and will be addressed in future work. For the present analysis, these measurements should therefore be regarded as upper limits on possible high-velocity Na\,I absorption.

\begin{figure}
    \centering
    \includegraphics[width=1.0\linewidth]{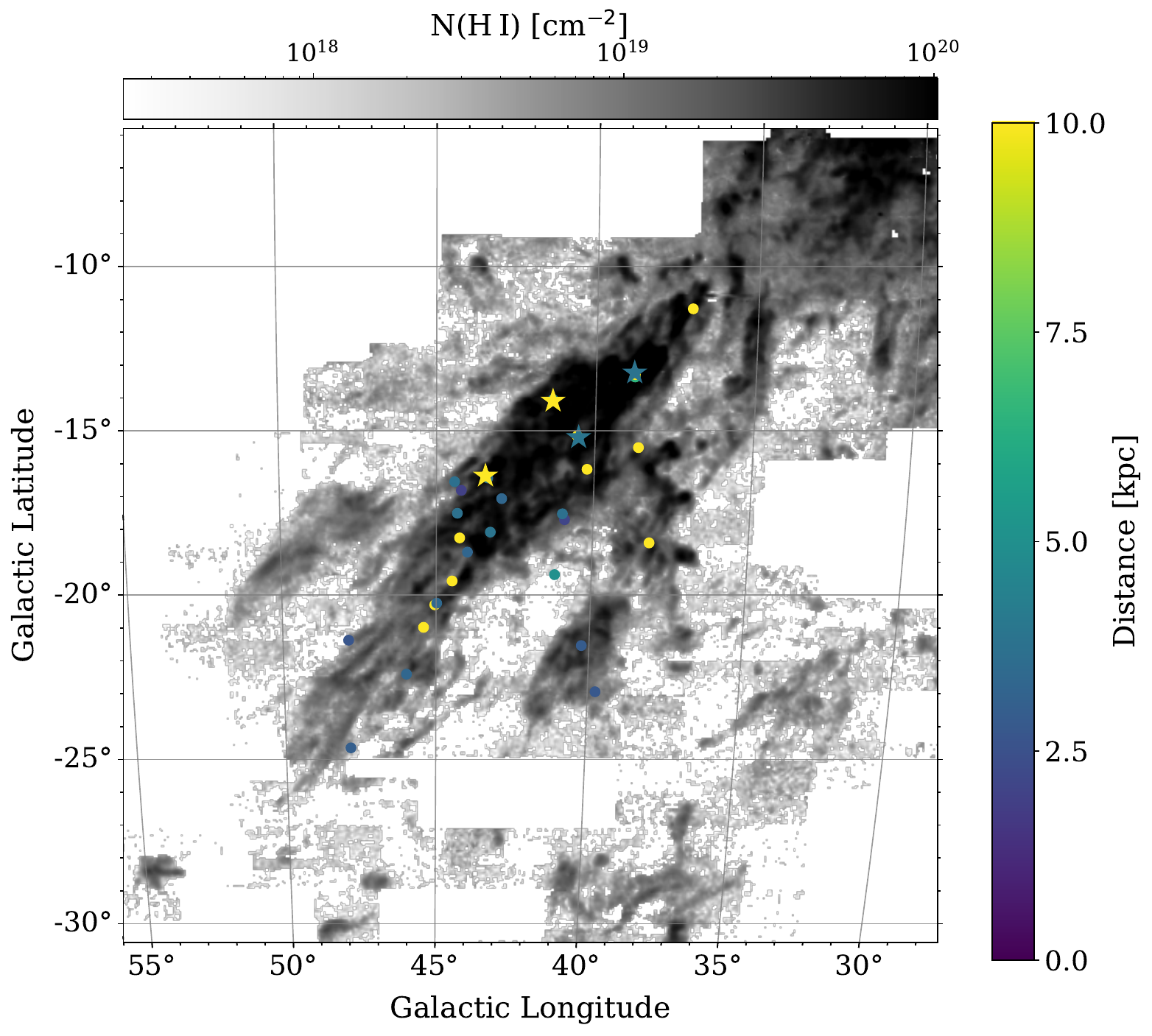}
    \caption{Positions of the 33 ARCES targets overlaid on the GBT 21-cm H\,I map of the Smith Cloud. Target markers are color-coded by Gaia distance. The four targets with $>3\sigma$ detections in Na\,I D$_2$ absorption are highlighted with star symbols.}

    \label{fig:echelle_HI}
\end{figure}

% Cursorily, this relatively low covering fraction of high velocity Na\,I detected in high-resolution spectra may seem to contrast with the signal identified in the stacked BOSS sample or, at least, raises questions about the structure of cold gas in the Smith Cloud. If the SC possesses a low Na\,I covering fraction, then it is plausible that we may miss the Na\,I absorption from the SC that we detect in our SDSS sample. Indeed, we find that the number of SC region Na\,I absorbers from our BOSS sample with equivalent widths that significantly exceed those from the control sample ($>2\sigma$) are consistent with such a low covering fraction ($f_c = 5-28\%$).  

Cursorily, this relatively low covering fraction of high-velocity Na\,I detected in high-resolution spectra may seem to contrast with the signal identified in the stacked BOSS sample or, at least, raise questions about the structure of cold gas in the Smith Cloud. If the Smith Cloud possesses a low Na\,I covering fraction, it is plausible that individual sightlines may miss the Na\,I absorption that is nevertheless recovered statistically in our SDSS sample. We quantify the Na\,I covering fraction, $f_c$, as the fraction of individual sightlines whose Na\,I equivalent-width excess exceeds $2\sigma$ relative to the median equivalent width of the corresponding distance-matched control bin. Using this definition, we measure a covering fraction of $f_c = 27^{+4}_{-3}\%$ toward the Smith Cloud region, consistent with a patchy distribution of cold, Na\,I--bearing gas. Given the uncertainty in the Smith Cloud distance, this covering fraction should be interpreted as an empirical measure of the incidence of excess Na\,I absorption along these sightlines rather than a direct constraint on the cloud’s internal structure.

The minimum angular separation between ARCES and BOSS targets is 1', and the ARCES sample was chosen to coincide with regions of strong H I emission, unlike the more randomly distributed BOSS targets. Given the large Na\,I/H I variations across small angular scales that have been observed in high-velocity gas \citep{Wakker+2001}, such variations across the SC are to be expected.

\subsection{Validation of Stellar Distances}
In this study, we adopt Gaia-based distances from \citet[][CBJ]{BailerJones+2021} because they provide uniform, all-sky coverage and well-characterized posteriors for both geometric and photogeometric estimates. The CBJ distances adopt a direction-dependent prior built from the GeDR3 mock (a Besançon-like model of galactic structure with PARSEC stellar evolutionary tracks; \citealt{Rybizki+2020}). This prior reflects the Gaia selection function and is therefore dominated by disk/bulge stars, with halo populations comparatively under-represented. In the low-S/N-parallax limit, the CBJ posterior reverts to the prior, which can pull true halo stars to smaller inferred distances. 

To quantify the impact of Gaia-based distances on our sample, we compare the CBJ catalog to alternative estimates. First, MINESweeper derives spectrophotometric distances by jointly fitting a subregion of the BOSS spectrum containing the Balmer H$\beta$ and the Mg\,I b triplet, broadband photometry, and parallaxes within a Bayesian framework constrained by MIST isochrones \citep{Cargile+2020,Chandra+2025}. Second, StarHorse derives spectrophotometric distances by combining Gaia parallaxes with multi-band photometry and spectroscopic stellar parameters ($T_{\rm eff}$, $\log g$, [M/H]) from several large surveys including APOGEE and BOSS, within a Bayesian framework constrained by PARSEC isochrones \citep{Queiroz+2018,Queiroz+2023}. Both of these approaches avoid the strong prior dependence of CBJ, yielding more reliable distances for distant and metal-poor halo populations. 

\begin{figure}
    \centering
    \includegraphics[width=0.9\linewidth]{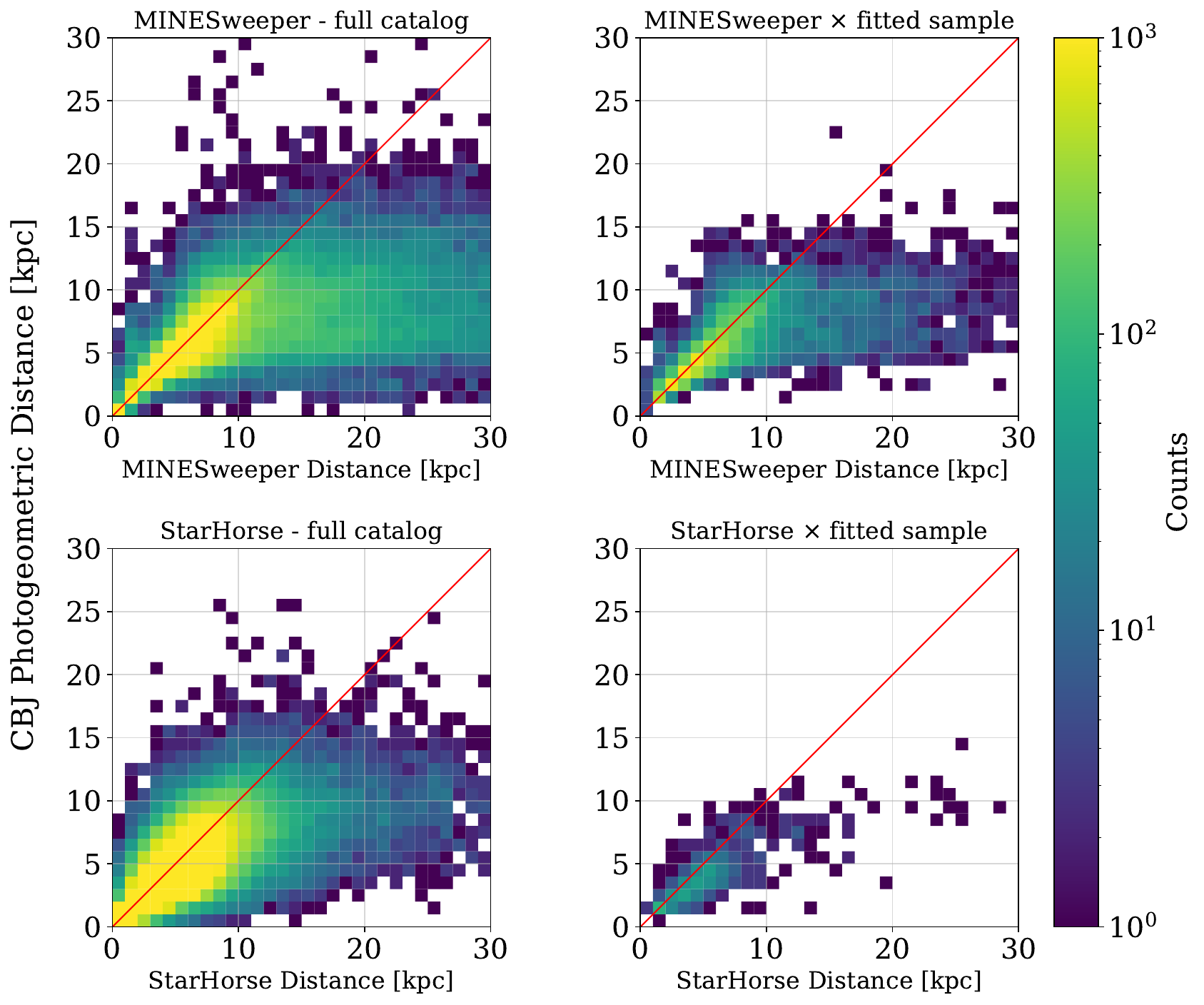}
    \caption{Comparison of Gaia-based photogeometric distances with MINESweeper (top) and StarHorse (bottom). The left panels show the full samples from each catalog, while the right panels contain stars cross-matched to our final fitted catalog. The solid red line marks the one-to-one relation. The two distance estimates agree well within 10 kpc, but at larger MINESweeper/StarHorse distances, the Gaia-based values lag behind.}
    \label{fig:distance_comparisons}
\end{figure}

We compare the CBJ distance to the MINESweeper and StarHorse estimates in Figure \ref{fig:distance_comparisons}. Both the full sample of MINESweeper and StarHorse as well as samples cross-matched to our catalog are compared. As expected, within 10 kpc, both catalogs are well correlated with the CBJ distances before leveling off at greater distances due to the prior constraints on the CBJ distances. As distance inconsistencies are mostly relevant beyond 10 kpc, the observed excess at $>$1 kpc in Na\,I is insensitive to this effect, but we note that inferences based on the CBJ catalog at $\gtrsim$10 kpc must account for potential prior-driven systematics in the CBJ distances.

\subsection{Foreground ISM, Dust Content, and Other Considerations}

The line of sight toward the Smith Cloud intersects a substantial amount of foreground interstellar material, including both atomic and molecular gas. From the GBT 21 cm data, the total integrated H\,I column density within a $10^\circ$ radius of the SC is $6.09 \times 10^{25}$ cm$^{-2}$, of which $1.93 \times 10^{24}$ cm$^{-2}$, approximately 3\%, lies within the SC velocity range (65–150 km\,s$^{-1}$ LSR). One prominent foreground component is a branch of the low-velocity Aquila molecular cloud complex, located at a distance of 200--400 pc \citep{Su+2020}, i.e., at closer distance than our observed Na\,I excess. Further casting doubt that Aquila is the source of excess Na\,I, it only subtends a small subregion of our SC analysis region. However, the distances to other foreground gas components remain uncertain, and their contributions to the observed excess in Na\,I absorption at $>$1 kpc cannot be disentangled from that of the Smith Cloud.

Another independent estimate of the dust content in the SC comes from \citet{Minter+2024}, who combined 21-cm H\,I and OH observations along three sightlines. They applied a standard N(H)/E(B–V) relation to derive the foreground and Smith Cloud reddening. Their analysis shows that low-velocity Aquila foreground gas accounts for nearly all of the FIR-derived extinction, implying E(B–V)$<$0.05 mag for the Smith Cloud. 

Another recent UV absorption study by \citet{Vazquez+2025} measured [Si/S] $\simeq -0.72$ along Smith Cloud sightlines. Because sulfur is a volatile element that is largely undepleted in diffuse gas, while silicon is more readily incorporated into dust grains, a subsolar [Si/S] ratio indicates that a fraction $(1-10^{[\mathrm{Si/S}]})$ of the silicon is locked into dust \citep{SavageSembach+1996,Jenkins+2009}. Converting this missing silicon mass into a dust mass relative to hydrogen following Equation~7 of \citet{Vazquez+2025} yields a dust-to-gas ratio of order $10^{-4}$--$10^{-3}$. This level of depletion corresponds to $E(B-V)$ of only a few $\times$10$^{-2}$ mag at most, consistent with our non-detection of enhanced reddening. Thus, the UV depletion evidence confirms a low but non-zero dust content in the Smith Cloud that remains below FIR detection thresholds yet agrees with our extinction measurements. Similar low dust-to-gas ratios have been reported for other high-velocity clouds, such as Complex~C \citep{Shull+2024,Fox+2023}, suggesting that the dust properties of the Smith Cloud are consistent with those of the broader HVC population.

\section{Conclusions} 
\label{sec:Conclusion}

We have leveraged a large sample of SDSS-V BOSS stellar spectra to probe Na\,I, Ca\,II, and dust extinction toward the Smith Cloud (SC), comparing with a same-latitude control field. Our main findings are as follows:

\begin{enumerate}
\item\textbf{All-sky absorption and extinction structure:} Across galactic latitudes and distance, our Na\,I and $A_V$ measurements recover expected structure and show clear distance trends: at low $|b|$, both rise steeply within the first few kpc; at high $|b|$, Na\,I plateaus by $\sim$4 kpc once the midplane is exited, whereas low-$|b|$ sightlines continue to increase beyond 4 kpc. The $A_V$ profiles likewise show a strong early rise at low $|b|$ and a more gradual increase at high $|b|$, consistent with dust and cold neutral gas concentrated toward the plane. The residual Ca\,II K map reveals a systematic offset between APO (north) and LCO (south) measurements, which we regard as an instrumental effect and aim to resolve in future work.

\item\textbf{Na\,I absorption enhancement in the direction of the SC:} The median residual Na\,I equivalent widths toward stars within $10^\circ$ of the SC show a clear excess at $d\gtrsim2\,$kpc compared to the control sample, consistent with cold gas in the SC being detected in low-resolution spectra. If this enhancement is due to the SC, it would imply a distance of 1--2 kpc, significantly closer than those of previous studies. 

%item\textbf{Low dust-to-gas ratio:} Joint fits of $A_V$ and Na\,I with both low- and high-velocity H\,I components show that high velocity gas within the vicinity of the Smith Cloud contributes at most a weak additional signal beyond the foreground ISM. The high-velocity term is marginally significant in the $A_V$ fit and not significant in the Na\,I fit, indicating that the dust-to-gas ratio in the SC is substantially lower than in the low-velocity ISM. This conclusion is consistent with previous studies that found little extinction associated with the SC \citep{Minter+2024} and large scatter in Na\,I/H\,I in high velocity gas \citep{Wakker+2001, Bekhti+2012}.

\item\textbf{Low dust-to-gas ratio:} 
We independently fit $A_V$ and Na\,I equivalent width as functions of both low- and high-velocity H\,I column density. In both cases, the low-velocity component associated with the foreground ISM dominates the signal. The high-velocity H\,I term associated with the Smith Cloud is marginally significant in the $A_V$ fit and Na\,I fit, indicating that the Smith Cloud contributes little dust and little cold Na-bearing gas per unit N(H\,I) compared to the low-velocity ISM. Although $A_V$ increases for stars at distances of 4--8 kpc, this behavior is equally present in the same-latitude control sample and therefore reflects foreground extinction unrelated to the Smith Cloud. This interpretation is consistent with previous studies that found minimal extinction associated with the Smith Cloud \citep{Minter+2024} and large scatter in Na\,I/H\,I in high-velocity gas \citep{Wakker+2001, Bekhti+2012}.

\item\textbf{Implications for SC structure:} The low covering fraction of Na\,I in high-resolution spectra along individual sightlines may reflect a low covering fraction or small-scale patchiness of cold and dusty clumps in the SC. This, together with the low overall extinction, suggests the SC has an inhomogeneous, clumpy morphology, with dense, colder pockets of gas embedded in a more diffuse envelope.
\end{enumerate}

The results of this proof-of-concept study establish SDSS-V BOSS stellar tomography as a powerful new tool for constraining distances, absorption structure, and dust content in HVCs across the Milky Way halo. Future works will further leverage the 3D absorption maps produced by this study to characterize the structure of inflows and outflows in the Galactic halo.

\begin{acknowledgments}
Funding for the Sloan Digital Sky Survey V has been provided by the Alfred P. Sloan Foundation, the Heising-Simons Foundation, the National Science Foundation, and the Participating Institutions. SDSS acknowledges support and resources from the Center for High-Performance Computing at the University of Utah. SDSS telescopes are located at Apache Point Observatory, funded by the Astrophysical Research Consortium and operated by New Mexico State University, and at Las Campanas Observatory, operated by the Carnegie Institution for Science. The SDSS web site is \url{www.sdss.org}.

SDSS is managed by the Astrophysical Research Consortium for the Participating Institutions of the SDSS Collaboration, including Caltech, The Carnegie Institution for Science, Chilean National Time Allocation Committee (CNTAC) ratified researchers, The Flatiron Institute, the Gotham Participation Group, Harvard University, Heidelberg University, The Johns Hopkins University, L'Ecole polytechnique f\'{e}d\'{e}rale de Lausanne (EPFL), Leibniz-Institut f\"{u}r Astrophysik Potsdam (AIP), Max-Planck-Institut f\"{u}r Astronomie (MPIA Heidelberg), Max-Planck-Institut f\"{u}r Extraterrestrische Physik (MPE), Nanjing University, National Astronomical Observatories of China (NAOC), New Mexico State University, The Ohio State University, Pennsylvania State University, Smithsonian Astrophysical Observatory, Space Telescope Science Institute (STScI), the Stellar Astrophysics Participation Group, Universidad Nacional Aut\'{o}noma de M\'{e}xico, University of Arizona, University of Colorado Boulder, University of Illinois at Urbana-Champaign, University of Toronto, University of Utah, University of Virginia, Yale University, and Yunnan University.
\end{acknowledgments}


%\vspace{5mm}

\clearpage

\appendix

\startlongtable
\begin{deluxetable}{lcccccc}
\centering 
\tablecolumns{7}
%%\tabletypesize{\tiny}
\tablewidth{0pt}
\tablecaption{Wakker et al. (2008) Targets}
\tablehead{\colhead{} & \colhead{} & \colhead{} & \colhead{Reported Distance} & \colhead{Gaia Distance} & \colhead{Ca\,II K EW} & \colhead{}\\ \colhead{Object} & \colhead{RA} & \colhead{DEC} & \colhead{(kpc)} & \colhead{(kpc)} & \colhead{(\AA)} & \colhead{U,L} \\ \colhead{(1)} & \colhead{(2)} & \colhead{(3)} & \colhead{(4)} & \colhead{(5)} & \colhead{(6)} & \colhead{(7)}}
\label{tab:Wakker Sources}
\startdata
2MASS J195741.61-004009.7 & 299.423 & -0.669 & 0.7$\pm$0.6 & 3.039$^{+0.46}_{-0.359}$ & $<$43 & …\\
2MASS J195927.29+000822.3 & 299.864 & 0.14 & 1.1$\pm$0.5 & 1.869$^{+0.081}_{-0.082}$ & $<$13 & L\\
2MASS J195925.49-000519.1 & 299.856 & -0.089 & 1.5$\pm$0.2 & 2.298$^{+0.123}_{-0.108}$ & $<$28 & …\\
2MASS J195912.00-002645.3 & 299.8 & -0.446 & 1.8$\pm$0.3 & 1.561$^{+0.041}_{-0.033}$ & $<$15 & L\\
2MASS J195922.75+000300.0 & 299.845 & 0.05 & 9.0$\pm$3.7 & 2.378$^{+0.467}_{-0.307}$ & $<$124 & …\\
V1172 Aql & 300.596 & -0.536 & 10.5$\pm$1.4 & 6.286$^{+1.247}_{-1.142}$ & $<$11 & L\\
2MASS J195823.36-002719.0 & 299.597 & -0.455 & 13.1$\pm$3.2 & 2.735$^{+0.361}_{-0.434}$ & $<$68 & …\\
V1084 Aql & 300.931 & 0.894 & 14.5$\pm$1.3 & 5.735$^{+1.388}_{-1.08}$ & 51$\pm$3 & U
\enddata
\tablecomments{\\
(1) Object name as reported by \citet{Wakker+2008}. \\
(2–3) Right ascension and declination of the star. \\
(4) Stellar distance reported by \citet{Wakker+2008}. \\
(5) Stellar distance via Gaia reported by \citet{BailerJones+2021}. \\
(6) Ca\,II K equivalent width measured by \citet{Wakker+2008}. \\
(7) Whether the star was used to place an upper (U) or lower (L) limit on the HVC's distance.}

\end{deluxetable}

\begin{rotatetable*}              
\begin{deluxetable*}{crrrrrrr}
\tabletypesize{\scriptsize}      
\tablewidth{0pt}                 
\tablecaption{High Resolution Sample from Apache Point Observatory\label{tab:apo_targeted}}
\tablehead{\colhead{} & \colhead{RA} & \colhead{DEC} & \colhead{Gaia Distance} & \colhead{D$_1$ low} & \colhead{D$_2$ low} & \colhead{D$_1$ high} & \colhead{D$_2$ high}\\ \colhead{Gaia ID} & \colhead{(deg)} & \colhead{(deg)} & \colhead{(kpc)} & \colhead{(m\AA)} & \colhead{(m\AA)} & \colhead{(m\AA)} & \colhead{(m\AA)}\\ \colhead{(1)} & \colhead{(2)} & \colhead{(3)} & \colhead{(4)} & \colhead{(5)} & \colhead{(6)} & \colhead{(7)} & \colhead{(8)}}

\startdata              
4235479426276870528 & 297.33 & -0.936 & 12.067$^{+4.196}_{-1.781}$ & 177±20 & 241±20 & -3±18 & 12±18\\
4236879551253119232 & 299.788 & -0.291 & 13.305$^{+2.451}_{-2.822}$ & 161±7 & 168±14 & 19±9 & 5±9\\
4237312483957706496** & 299.203 & 0.88 & 16.684$^{+3.923}_{-4.253}$ & 415±8 & 456±8 & 99±16 & 210±16\\
4243541835799621504* & 302.175 & 1.566 & 12.441$^{+1.174}_{-1.102}$ & 222±10 & 317±10 & -13±16 & 60±15\\
4230700158458721792 & 306.344 & 0.973 & 13.85$^{+0.948}_{-1.056}$ & 174±23 & 180±28 & 12±16 & -11±16\\
4238179483237769088 & 294.765 & -1.528 & 12.834$^{+2.306}_{-2.476}$ & 43±53 & -55±49 & 5±61 & 25±52\\
4243537712630989440 & 302.2 & 1.474 & 3.634$^{+0.207}_{-0.253}$ & 71±11 & 106±14 & 0±12 & 6±11\\
4240291507631868672 & 299.167 & 0.9 & 4.519$^{+0.378}_{-0.397}$ & -3±5 & 12±5 & 1±6 & 3±6\\
4235480731946998656** & 297.287 & -0.895 & 3.822$^{+0.319}_{-0.195}$ & 177±6 & 215±7 & 77±8 & 93±8\\
4230886147722766208 & 306.271 & 0.948 & 3.367$^{+0.172}_{-0.184}$ & 66±27 & 31±39 & -10±11 & 13±12\\
4243541767080139776 & 302.192 & 1.56 & 5.566$^{+0.499}_{-0.456}$ & 109±23 & 50±44 & -3±16 & 9±16\\
4236877936345390848* & 299.804 & -0.351 & 3.897$^{+0.21}_{-0.196}$ & 115±17 & 100±20 & 1±20 & 72±20\\
4235467434728094336 & 297.396 & -0.971 & 6.465$^{+0.531}_{-0.537}$ & 63±11 & 36±13 & -6±12 & -4±14\\
4230545333477496448 & 305.454 & 0.874 & 14.808$^{+1.763}_{-1.514}$ & 60±20 & -46±20 & 15±23 & -10±30\\
4234391906194154112 & 295.625 & -3.888 & 15.041$^{+3.151}_{-2.912}$ & -43±25 & -121±31 & -17±30 & 29±27\\
4242630989195852672 & 299.225 & -2.112 & 12.636$^{+1.313}_{-1.348}$ & 111±14 & 40±15 & 10±17 & 11±21\\
4221729964647430528 & 304.203 & 1.321 & 16.125$^{+2.974}_{-3.029}$ & -78±22 & -219±25 & 12±27 & 2±30\\
4235983724152830848 & 301.604 & -3.83 & 14.824$^{+6.001}_{-4.399}$ & -353±38 & -536±40 & 34±63 & 7±62\\
4230734032864045696 & 300.53 & -1.053 & 16.24$^{+3.3}_{-2.175}$ & -39±19 & -233±20 & 65±26 & 5±25\\
4236735205988783488 & 307.108 & 0.932 & 3.308$^{+0.277}_{-0.18}$ & 115±16 & 131±19 & -3±17 & -3±15\\
4235841199959650688 & 302.542 & 0.792 & 3.663$^{+0.219}_{-0.208}$ & 182±10 & 167±12 & 2±12 & -11±12\\
4243732154388823168 & 302.062 & -1.058 & 3.642$^{+0.311}_{-0.279}$ & 121±12 & 213±12 & 5±11 & 19±11\\
4219684804295593344 & 302.787 & 2.297 & 2.881$^{+0.159}_{-0.145}$ & 95±5 & 143±6 & 2±8 & 18±8\\
4242473591531498752 & 305.283 & -3.552 & 3.86$^{+0.313}_{-0.243}$ & 113±16 & 167±17 & 15±19 & -7±18\\
4223872294332098304 & 303.596 & 0.592 & 4.964$^{+0.521}_{-0.428}$ & 44±30 & 103±28 & 26±38 & 7±38\\
4242546228017266560 & 303.787 & -1.763 & 3.271$^{+0.195}_{-0.144}$ & 119±15 & 98±17 & 15±17 & 7±16\\
4242894086009676800 & 304.463 & 0.901 & 3.699$^{+0.358}_{-0.258}$ & 44±24 & 71±28 & 24±27 & -3±28\\
4243675628321792384 & 303.583 & 1.751 & 1.94$^{+0.071}_{-0.087}$ & 153±6 & 170±7 & 23±8 & 16±8\\
4231414802360067328 & 302.917 & 1.991 & 3.298$^{+0.221}_{-0.166}$ & 158±5 & 192±5 & 1±6 & 7±6\\
4235823740909804160 & 302.917 & 1.991 & 2.06$^{+0.075}_{-0.076}$ & 214±7 & 218±7 & 24±9 & 6±8\\
4228629812422575104 & 308.604 & 0.676 & 2.993$^{+0.193}_{-0.156}$ & 123±7 & 149±7 & 0±9 & 2±9\\
4232525519560784512 & 302.187 & -1.204 & 2.659$^{+0.158}_{-0.156}$ & 74±10 & 124±17 & 4±10 & 5±10\\
4218952392113608064 & 305.283 & -3.552 & 2.711$^{+0.213}_{-0.15}$ & 139±15 & 186±13 & 13±17 & 11±17\\
\enddata  
\tablecomments{\\ (1) Gaia DR3 ID. * denotes a 3$\sigma$ detection in the D$_2$ high velocity Na\,I line, and ** denotes a 3$\sigma$ detection in both D$_2$ and D$_1$ transitions.\\
(2–3) Right ascension and declination of the star. \\
(4) Stellar distance via Gaia reported by \citet{BailerJones+2021}. \\
(5–6) Equivalent width measured for the Na\,I D$_1$/D$_2$ lines at low velocity (-20 -- 20 km s$^{-1}$ LSR), with 1$\sigma$ errors. \\
(7–8) Equivalent width measured for the Na\,I D$_1$/D$_2$ lines at high velocity (70 -- 150 km s$^{-1}$ LSR), with 1$\sigma$ errors.}

\end{deluxetable*}
\end{rotatetable*}

\bibliography{sample631}{}
\bibliographystyle{aasjournal}



\end{document}